\documentclass[12pt]{article}
\usepackage{amssymb, amsmath, bm, float, epsfig, color}
\usepackage{float}

\newcommand{\cred}[1]{{#1}}
\newcommand{\cblue}[1]{{#1}}

\begin{document}
\title{
\vspace{-20pt}
\begin{flushright}
\normalsize WU-HEP-17-03 \\
KIAS-P17014 \\*[55pt]
\end{flushright}
{\Large \bf Brane-localized masses in magnetic compactifications
\\*[20pt]}
}

\author{
Makoto~Ishida,$^1$\footnote{\cred{ncanis3@fuji.waseda.jp}} \,\, Kenji~Nishiwaki,$^2$\footnote{\cred{nishiken@kias.re.kr}} \, and \, Yoshiyuki~Tatsuta,$^1$\footnote{\cred{y\_tatsuta@akane.waseda.jp}}\\*[30pt]
$^1${\it \normalsize Department of Physics, Waseda University, Tokyo 169-8555, Japan}\\
$^2${\it \normalsize School of Physics, Korea Institute for Advanced Study~(KIAS), Seoul 02455, Republic of Korea}\\*[55pt]
}

\date{
\centerline{\small \bf Abstract}
\begin{minipage}{0.9\textwidth}
\medskip\medskip 
\small
{We discuss effects of the brane-localized mass terms on the fixed points of the toroidal {orbifold} $T^2/Z_2$ under the presence of background magnetic fluxes, where multiple lowest and higher-level Kaluza--Klein~(KK) modes are realized before introducing the localized masses in general.
Through the knowledge of linear algebra, we find that, in each KK level, one of or more than one of the degenerate KK modes {are almost inevitably} perturbed, when single or multiple brane-localized mass terms are introduced.
When the typical scale of the compactification is far above the electroweak scale or the TeV scale, we apply this mechanism {for uplifting} unwanted massless or light modes which are prone to appear {in models on} magnetized orbifolds.}
\end{minipage}}

\begin{titlepage}
\maketitle
\thispagestyle{empty}
\end{titlepage}

\tableofcontents

\section{Introduction}

The standard model (SM) of {particle physics} has been verified by the discovery of the final puzzle \cred{piece---i.e., the Higgs boson---in} 2012 \cite{Aad:2012tfa, Chatrchyan:2012ufa}.
It is well known that the SM can explain almost all of the phenomena around the electroweak scale ($\sim 100$ GeV) with great \cred{accuracy.}
However, {extensions} beyond the SM are required, due to several theoretical difficulties which appear {inevitably} in the SM, e.g., the flavor puzzle, the lack of a dark matter candidate, the {gauge} hierarchy problem\cred{,} and so forth.

Among such \cred{extensions} beyond the SM, extra dimensions have been studied from a phenomenological point of view.
Indeed, the geometry of the compactified {hidden directions} determines phenomenological properties in the four-dimensional (4D) low-energy effective theory (LEET).
For example, it is known that \cred{some {extra-dimensional} properties} such as \cred{Kaluza-Klein (KK) wave functions} {reflect} information on the {extra-dimensional} topologies.
In particular, {localization} of the lowest \cred{wave function}(s) among the KK-decomposed modes {significantly} affects the LEET {obtained} after dimensional reduction.
Indeed, many phenomenological models {accompanying} localization of the {KK-expanded} modes have been proposed and investigated.
For example, overlap integrations of {KK \cred{wave functions}} are used to realize a huge hierarchy in {Yukawa} coupling constants \cite{ArkaniHamed:1999dc}, where differences {in} the degrees of overlapping lead to the hierarchy in the eigenvalues of the Yukawa matrix.
Similarly, such an overlapping of {KK} \cred{wave functions} can \cred{provide} the \cred{Froggatt-Nielsen} mass matrix textures \cite{Gherghetta:2000qt}, their {Gaussian-extended} version \cite{Abe:2014vza}\cred{,} and so {on}.
\cblue{For such a reason}, \cred{model builders interested in extra dimensions are intrigued by the} localization of particle profiles in extra {directions} \cred{as a way} to realize a huge hierarchy in a natural way.

On the other hand, when we {address} orbifold compactifications, an interesting feature {is found, which is useful for concrete model constructions\cred{:}} the existence of orbifold fixed point(s).
{Model builders} have added {desirable} terms on the orbifold fixed points {to derive necessary structures and/or to conquer problems \cred{that are more difficult} in the bulk part of extra dimensions}.
For example, \cred{the authors of} Ref.\,\cite{Hall:2001pg} {pointed out that} the Yukawa couplings can be introduced on {the} $S_1/Z_2$ orbifold fixed points, although Yukawa interactions are prohibited on the bulk of $S_1/Z_2$.
There are {other} \cred{uses of} fixed points. 
In some papers, {model builders} have introduced brane-localized mass term(s) on the fixed points in order to {uplift dangerous massless (or very light) particles for consistent} model constructions.
In Refs.\,\cite{Dudas:2005vn, Sakamura:2016kqv}, brane-localized {mass terms on the fixed points of the} toroidal orbifold $T^2/Z_2$ {were} investigated and it {was declared} that a massless \cred{zero mode} can become massive via effects of the brane-localized mass.

Recently, several groups have eagerly studied systems {on magnetized backgrounds based on toroidal extra dimensions and their orbifolded versions} \cite{Buchmuller:2015eya, Buchmuller:2015jna, Buchmuller:2016dai, Buchmuller:2016bgt, Buchmuller:2016gib, Buchmuller:2017vho, Abe:2013bca, Abe:2014noa, Abe:2015yva, Fujimoto:2016zjs, Kobayashi:2016qag, Abe:2016jsb}.\footnote{See also Refs.\,\cite{Fujimoto:2013xha, Higaki:2016ydn}.}
This is because {magnetic fluxes} play important roles in constructing phenomenological models \cite{Cremades:2004wa}.
Indeed, {the presence of magnetic fluxes leads} to the multiplicity of the lowest KK modes, {where such {an} emergence of the multiple lowest modes is considered as a spontaneous generation of a family replication in LEET}.
Also, specific configurations of {the magnetic fluxes penetrating extra dimensions} break supersymmetry, \cred{(see e.g., Ref.\,\cite{Abe:2014vza})}.

{In this paper, we examine the situation \cred{where the two phenomenology fascinating ideas—namely, a} magnetized background and \cred{a} mass term localized on \cred{an orbifold fixed point---are} taken into account simultaneously.
Our major motivation \cred{for focusing} on this configuration is as follows.
A possible problem in realizing family structures \cred{when using} the magnetic fluxes is that {extra massless modes} emerge in \cred{some} concrete models.
\cred{Introducing} brane-localized mass terms may help such situations by making some of the {light} particles decoupled, which can be expected.
Also, after the perturbation by the insertion of mass terms on fixed points, particle profiles are changed through \cred{the rediagonalization} of a perturbed {KK} mass matrix, where {some} part of \cred{the} mass spectrum may be unchanged.}


This paper is organized as follows.
In Sec.\,\cblue{\ref{sec:review}}, we briefly review the KK decompositions for the six-dimensional (6D) scalar and spinor fields {on \cred{a} magnetized} {two-dimensional torus}, and show that the {KK-expanding} \cred{wave functions} {are described} by the Jacobi theta function and the Hermite polynomials {on the basic magnetic background}.
\cred{Then, we focus on} the $Z_2$-orbifolded situation of \cred{a} two-dimensional torus with magnetic fluxes, {where important properties of the \cred{$Z_2$ eigenmodes} are shown}.
In Sec.\,\cblue{\ref{sec:mass_perturbation}}, we investigate effects on the KK mass spectra {after taking care of effects \cred{from} brane-localized mass(es)}, where \cred{the} forms of eigenvalues and eigenvectors after the perturbation are investigated \cred{theoretically.}
\cred{Subsequently,} in Sec.\,\cblue{\ref{sec:wavefunction}} we directly explore deformations of the profile of KK particles (in the correct mass eigenbases under the brane-localized mass terms) through numerical calculations.
In Sec.\,\cblue{\ref{sec:cutoff}}, we comment on \cred{the cutoff dependence of the} KK mass eigenvalues.
Section\,\cblue{\ref{sec:conclusion}} is devoted to \cred{a} conclusion and discussion.
{In \cred{Appendix}\,\ref{app:notation}, we summarize our notation for 6D gamma matrices.
In \cred{Appendix}\,\ref{app:multiple_massterm}, we provide a discussion on the situation with multiple brane-localized mass terms.}

\section{Brief review \cred{of} bulk \cred{wave functions}}
\label{sec:review}

In this section, we briefly review the \cred{wave functions} of {KK} modes on \cred{a} magnetized background, based on Refs.\,\cite{Cremades:2004wa, Abe:2008fi, Hamada:2012wj, Abe:2013bca}.

\subsection{Flux background}

We consider {the two actions of the 6D gauge theory on the 4D Minkowski spacetime times two-dimensional torus $T^2$ with a 6D Weyl spinor ($\Psi$) or a 6D complex scalar ($\Phi$), e.g.,
\begin{align}
S_{\rm Weyl} &= \int d^4 x \int_{T^2} d^2z \, \{ i \bar \Psi \Gamma^M D_M \Psi \}, \\
S_{\rm scalar}   &= \int d^4 x \int_{T^2} d^2z \, \left\{ - |D_M \Phi|^2 \right\},
\end{align}
where} \cred{the} index $M$ runs over $\mu \, (= 0, 1, 2, 3), 5, 6$ and {$\Gamma^M$ denotes the 6D gamma matrices (see \cred{Appendix \ref{app:notation} for details} of our notation).}
$D_M \equiv \partial_M - iq A_M$ denotes a covariant derivative.
Here, $q$ denotes a $U(1)$ charge. 
In the above action, we define a (dimensionless) complex coordinate $z \equiv (y_5 + \tau y_6)/2\pi R$ with two Cartesian coordinates $y_5$ and $y_6$ and $\tau \in \mathbb{C}$ to express two extra space directions.
{Also}, $R$ denotes a compactification radius of $T^2$ and is associated with a \cred{compactification} scale $M_C$, i.e.,  $M_C \sim 1/R$.
{In} the complex coordinate, the toroidal periodic condition is expressed as $z \sim z + 1 \sim z + \tau$.

In the {six-dimensional} action, we assume that the vector potential $A_M$ possesses \cred{a} (classical) {nontrivial} flux background $b = \int_{T^2} F$ \cred{with} field strength $F = (ib/2 {\rm Im} \, \tau) dz \wedge d \bar z$:
\begin{gather}
A^{(b)} (z) = \frac{b}{2 {\rm Im} \, \tau} {\rm Im} \, \bar z dz.
\end{gather}
The consistency condition under \cred{contractible} loops, e.g., $z \to z+1 \to z+1+\tau \to z+\tau \to z$, provides the Dirac charge quantization,
\begin{gather}
\frac{qb}{2\pi} = M \in \mathbb{Z}.
\end{gather}
Indeed, the magnetic flux plays an important role in the context of {higher-dimensional} gauge theory.
For example, it \cred{was} found in Ref.\,\cite{Cremades:2004wa} that the flux background can provide the multiplicity of KK-{expanding} \cred{wave functions} and their {localized profiles}, as we will see \cred{below}.

\subsection{KK modes on $T^2$ with magnetic fluxes}

{Next}, we briefly review KK-mode \cred{wave functions} of {6D Weyl spinor and scalar fields, denoted by $\Psi$ and $\Phi$}, on \cred{a} {magnetized} $T^2$, based on \cblue{Refs.}\,\cite{Cremades:2004wa, Hamada:2012wj}.
First, we decompose them as
\begin{align}
\Psi (x^\mu, z) &= \sum_n \chi_n (x^\mu) \otimes \psi_n (z), \label{KKexpand1}\\
\Phi (x^\mu, z) &= \sum_n \varphi _n(x^\mu) \otimes \phi_n (z), \label{KKexpand2}
\end{align}
{where the integer $n = 0,1,2,\cdots$ discriminates values of KK levels.}
For later convenience, {we adopt the above notation where} a \cred{two-dimensional(2D)} spinor $\psi_n$ carries a 2D chirality distinguished by $\pm$ as $\psi_n = (\psi_{+, n}, \, \psi_{-, n})^\text{T}$. 
The KK modes of the spinor in Eq.\,\eqref{KKexpand1} are {designated as} eigenstates of the covariant derivative {$D \equiv \left[ \partial_{\bar{z}} + \pi M z/(2{\rm Im} \, \tau) \right]/(\pi R)$} with $\partial_{\bar{z}} = \pi R (\partial_{y_5} + \bar\tau \partial_{y_6})$ as 
\begin{align}
\begin{pmatrix}
D^\dag D & 0\\
0 & DD^\dag
\end{pmatrix}
\begin{pmatrix}
\psi_{+, n}\\ \psi_{-, n}
\end{pmatrix}
= m_n^2 
\begin{pmatrix}
\psi_{+, n}\\ \psi_{-, n}
\end{pmatrix}
,
\label{modeeq1}
\end{align}
while those of the scalar in Eq.\,\eqref{KKexpand2} are eigenstates of the Laplace operator $\Delta \equiv \{D^\dag, D\}/2$ as
\begin{gather}
\Delta \phi_n = m_n^2 \phi_n.
\label{modeeq2}
\end{gather}

For simplicity, we \cred{choose a simple} complex structure parameter, i.e., $\tau=i$.
Also, we focus on \cred{the} case of positive magnetic fluxes $M>0$.
{Because} it is straightforward to apply the following discussions to {nontrivial} values of $\tau$ and negative fluxes, we will not \cred{address} such a {case} on this paper.

The form of the eigenstates of the KK modes {is} shown analytically by the Jacobi theta function and the Hermite polynomials.
First, we focus on (massless) zero-mode \cred{wave functions} for $\psi_{+, n}$.
The zero-mode \cred{wave functions} are multiply degenerate and given as
\begin{gather}
\psi^j_{+, 0}(z) = {{\cal N}} e^{\pi i Mz {\rm Im} \, z} \, \vartheta
\begin{bmatrix}
j/M \\[3pt] 0
\end{bmatrix}
(Mz, Mi),
		\label{eq:zero_mode_form}
\end{gather}
where the Jacobi theta function is defined by
\begin{gather}
\vartheta
\begin{bmatrix}
a\\[3pt] b
\end{bmatrix}
(\nu, \tau) =
\sum_{\ell \in \mathbb{Z}} e^{\pi i (a+ \ell)^2 \tau + 2\pi i (a + \ell) (\nu +b)},
\end{gather}
where $a$ and $b$ are real parameters, and $\nu$ and $\tau$ take complex values with ${\rm Im} \, \tau> 0$.
In the above expression, the number of degenerate \cred{zero modes} is determined by the magnitude of \cred{the} magnetic fluxes, i.e., $j=0, 1,..., M-1$ {(with modulo $M$)}.
{The} normalization constant is calculated as ${{\cal N}} = (2M/{\cal A}^2)^{1/4}$\cred{, where the area of the} torus ${\cal A} = (2\pi R)^2$, {which is independent of $j$ on $T^2$.}
On the other hand, massive KK-mode {\cred{wave functions} are} given as
\begin{gather}
\psi_{+, n}^j(z) = \frac{{{\cal N}}}{\sqrt{2^n n!}} e^{\pi i M z {\rm Im} \, z}
 \sum_{\ell \in \mathbb{Z}} e^{ -\pi M (\frac{j}{M} + \ell )^2 + 2\pi i Mz (\frac{j}{M} + \ell )}
H_n \bigl( \sqrt{2\pi M} \left(\tfrac{j}{M} + \ell + {\rm Im} \, z \right) \bigr),
		\label{eq:KK_mode_form}
\end{gather}
with the Hermite polynomials
\begin{gather}
H_n(x) = (-1)^n e^{x^2} \frac{d^n}{dx^n} e^{-x^2}{.}
\end{gather}
{We note that the form in Eq.\,(\ref{eq:zero_mode_form}) is a specific case {($n=0$)} of that in Eq.\,(\ref{eq:KK_mode_form}).}
As shown in Refs.\,\cite{Cremades:2004wa, Hamada:2012wj}, the {squared} KK mass eigenvalue is given as
\begin{gather}
m_n^2= \frac{4 \pi M}{\cal A} n,
	\label{eq:KKmass_fermion}
\end{gather}
for {$n = 0,1,2,\cdots$}, {which is independent of the index $j$.}

\cred{Next, we address} the case of $\psi_{-,n}$.  
Indeed, since \cred{nonvanishing} magnetic fluxes cause a chirality projection for massless \cred{zero modes} of $\psi_n$, \cred{the zero modes} $\psi_{-, 0}$ are not normalizable for $M>0$.
For $n \geq 1$, \cred{wave functions} of the KK modes are similarly written as $\psi^j_{-, n} = D \psi^j_{+, n}/m_n$. 
Note that the multiplicity of $\psi^j_{-, n}$ is the same as that of $\psi^j_{+,n}$ {when $n \geq 1$}.

{The} case of the 6D scalar field is {treated} similarly to the {case} of the 6D spinor {which we discussed}. 
The {set of} \cred{wave functions} of the scalar field is exactly the same as {that} of the spinor, i.e.,
\begin{gather}
\phi_{n}^j(z) = \frac{{{\cal N}}}{\sqrt{2^n n!}} e^{\pi i M z {\rm Im} \, z}
 \sum_{\ell \in \mathbb{Z}} e^{ -\pi M (\frac{j}{M} + \ell )^2 + 2\pi i Mz (\frac{j}{M} + \ell )}
H_n \bigl( \sqrt{2\pi M} \left(\tfrac{j}{M} + \ell + {\rm Im} \, z \right) \bigr),
\end{gather}
where $n \geq 0$ {and $j = 0,1,\cdots,M-1$ (with modulo $M$)}.
{An important} difference between the scalar and spinor fields {is found} in their mass eigenvalues.
The KK mass {spectrum of} the scalar {is} given as
\begin{gather}
m_n^2 = \frac{2\pi M}{\cal A} (2n+1),
	\label{eq:KKmass_scalar}
\end{gather}
{which} implies that the lowest KK modes of the scalar are {massive}.
\footnote{{If one tries to embed the toroidal compactification with fluxes into the superstring/supergravity theories, it is plausible that the above charged (fundamental) scalar field may \cred{consist} of some \cred{higher-dimensional} gauge fields as a possibility \cred{for} the UV completion.
\cred{Although} a derivation of the scalar is often difficult, \cred{in this paper} we analyze the scalar spectrum from the \cred{field-theoretical} point of \cred{view.}}}

\subsection{KK modes on $T^2/Z_2$ with magnetic fluxes}

Now, we are ready to \cred{address the wave functions} on $T^2/Z_2$ with fluxes.
In addition to {toroidal conditions on the fields}, we introduce an additional identification {in} the 2D space. 
In general, for $N=2, 3, 4\cred{,}$ and $6$, the $T^2/Z_N$ orbifold is {defined by identifications under the twist},
\begin{gather}
z \sim e^{2 \pi i /N} z.
\end{gather}
It {was concretely pointed out} in Refs.\,\cite{Abe:2008fi, Abe:2013bca} that the magnetic fluxes can coexist with the twist identification, and also that some parts of the KK-expanded modes are projected out.
Accordingly, the multiplicity of the KK modes {is} changed and hence magnetized toroidal orbifolds can be an interesting framework for phenomenological model \cred{building}.\footnote{
Another motivation for considering magnetized orbifolds is to realize the \cred{$CP$} violation in the quark sector LEET \cred{via} higher-dimensional supersymmetric \cred{Yang-Mills} theories (see {Ref.}\,\cite{Kobayashi:2016qag}).
}

In this paper, we restrict ourselves to the $Z_2$ twisted orbifold as an illustration. 
Also, we assume that {(discretized)} Wilson line and \cred{Scherk-Schwarz} twisting phases are all vanishing.
Since an extension to {the cases} with {such} \cred{nonvanishing} twisting phases can be done straightforwardly by means of the operator formalism \cite{Abe:2014noa}, we {\cred{do not} address such generalized situations}.
Under the above twist identification, we {construct} the $Z_2$ eigenstates {of} the KK modes {which should obey the boundary conditions around $z=0$}:
\begin{align}
\psi_{T^2/Z_2 \, \pm, n}(-z) &= \pm \eta \psi_{T^2/Z_2 \, \pm, n}(z),
\end{align}
where $\eta$ denotes the $Z_2$ \cred{parity} $\eta = \pm 1$.

It {was} pointed out in Refs.\,\cite{Abe:2008fi, Abe:2013bca} that the physical eigenstates $\psi_{T^2/Z_2 \, \pm, n}$ on $T^2/Z_2$ are easily obtained {as}
\begin{align}
\psi^j_{T^2/Z_2 \, +, 0}(z) &= \frac1{\sqrt 2} (\psi^j_{+, 0} (z) + \eta \psi^{j}_{+, 0}(-z)) \notag\\
&= \frac1{\sqrt 2} (\psi^j_{+, 0} (z) + \eta \psi^{M-j}_{+, 0}(z)),
\label{eigenstate1}
\end{align}
where we \cred{used the} important property $\psi^j_{+, 0}(-z) = \psi^{M-j}_{+, 0}(z)$. 
This expression is just a formal solution of the zero-mode equations in Eqs.\,\eqref{modeeq1} and \eqref{modeeq2}.
For \cred{an arbitrary number} of {quantized} fluxes, the \cred{number} of independent zero-mode \cred{wave functions is} counted as shown in \cred{Table} \ref{numofneven}.

Next, \cred{normalizable wave functions} of the excited KK modes ($n \geq 1$) are similarly written as
\begin{gather}
\psi^j_{T^2/Z_2 \, \pm, n}(z) = \frac1{\sqrt 2} (\psi^j_{\pm, {n}} (z) \pm \eta (-1)^n \psi^{M-j}_{\pm, {n}}(z)),
\label{eigenstate2}
\end{gather}
where we use a similar formula for the KK modes on $T^2$: $\psi^j_{\pm, n}(-z) = (-1)^n \psi^{M-j}_{\pm, n}(z)$. 
The \cred{eigen--wave functions} in Eqs.\,\eqref{eigenstate1} and \eqref{eigenstate2} {keep} the same mass spectrum as {those} on $T^2$, i.e.,
\begin{gather}
\begin{pmatrix}
D^\dag D & 0 \\
0 & DD^\dag
\end{pmatrix}
\begin{pmatrix}
\psi_{T^2/Z_2 \, +, n}\\
\psi_{T^2/Z_2 \, -, n}
\end{pmatrix}
=m_n^2
\begin{pmatrix}
\psi_{T^2/Z_2 \, +, n}\\
\psi_{T^2/Z_2 \, -, n}
\end{pmatrix},\\
m_n^2 = \frac{4\pi M}{\cal A} n,
\end{gather}
except for $\psi_{-, 0}$, {which has no consistent solution when $M>0$ on $T^2$.}
These expressions are also formal solutions and the \cred{number} of independent {physical modes \cred{is} calculated}. 
In Tabs.\,\ref{numofneven} and \ref{numofnodd}, the \cred{number} of independent KK \cred{wave functions} \cred{is} shown. 
{Here, we mention \cred{the} {ranges} of the index $j$ after the $Z_2$ orbifolding.
The index $j$ starts from zero or one in a $Z_2$-even \cred{or} $Z_2$-odd case, respectively, since the $j=0$ component apparently vanishes in the latter case.}
{Also, to avoid double counting, when the number of independent physical modes is $n_{\text{mode}}$, the first $n_{\text{mode}}$ values of $j$ are taken as individual degrees of freedom.}

\begin{table}[t]
\centering 
\begin{tabular}{|c|cccccccccc|cc|} \hline
$M$ & $0$ & $1$ & $2$ & $3$ & $4$ & $5$ & $6$ & $7$ & $8$ & $9$ & $2k$ & $2k+1$ \\ \hline
$\eta=+1$ & $1$ & $1$ & $2$ & $2$ & $3$ & $3$ & $4$ & $4$ & $5$ & $5$ & $k+1$ & $k+1$\\
$\eta=-1$ & $0$ & $0$ & $0$ & $1$ & $1$ & $2$ & $2$ & $3$ & $3$ & $4$ & $k-1$ & $k$ \\ \hline
\end{tabular}
\caption{The \cred{number} of independent KK \cred{wave functions} for even $n$ on $T^2/Z_2$ with fluxes.
{The general forms are valid for $M \geq 1$.}}
\label{numofneven}
\vspace{15pt}
\begin{tabular}{|c|cccccccccc|cc|} \hline
$M$ & $0$ & $1$ & $2$ & $3$ & $4$ & $5$ & $6$ & $7$ & $8$ & $9$ & $2k$ & $2k+1$ \\ \hline
$\eta=+1$ & $0$ & $0$ & $0$ & $1$ & $1$ & $2$ & $2$ & $3$ & $3$ & $4$ & $k-1$ & $k$ \\
$\eta=-1$ & $1$ & $1$ & $2$ & $2$ & $3$ & $3$ & $4$ & $4$ & $5$ & $5$ & $k+1$ & $k+1$\\ \hline
\end{tabular}
\caption{The \cred{number} of independent KK \cred{wave functions} for odd $n$ on $T^2/Z_2$ with fluxes.
{The general forms are valid for $M \geq 1$.}}
\label{numofnodd}
\end{table}

Before closing this section, it is important to discuss {the orthonormal condition} for the physical eigenstates of the KK modes on $T^2/Z_2$.
Let us consider an overlap integral of physical states on $T^2/Z_2$,
\begin{gather}
\int_{T^2} d^2z \, \psi^j_{T^2/Z_2 \, \pm, n} \bigl( \psi^{j'}_{T^2/Z_2 \, \pm, n'} \bigr)^\dag = \delta_{n, n'} \bigl( \delta_{j, j'}+ \eta (-1)^n \delta_{j+j', M} \bigr),
\end{gather}
where the Kronecker delta appearing in this relation should be interpreted {as} that with modulo $M$.
It is easy to \cred{see} that the second term \cred{on} the {right-hand} side provides a nonzero contribution only for $j+j'=M$, {which is rephrased as} $j = j' = M/2 ~~ ({\rm mod} \, M/2)$ for $\eta = +1$. 
Thus, by redefining a normalization constant as
\begin{align}
{{\cal N}} = (2M/{\cal A}^2)^{1/4} \implies {\cal N}_j = (2M/{\cal A}^2)^{1/4}/\sqrt{1+\delta_{j, M/2}},
\end{align}
the \cred{wave functions are} normalized and \cred{orthogonal to} each other, such \cred{as}
\begin{align}
\int_{T^2} d^2z \, \psi^j_{T^2/Z_2 \, \pm, n} \bigl( \psi^{j'}_{T^2/Z_2 \, \pm, n'} \bigr)^\dag = \delta_{n, n'} \delta_{j, j'}.
\end{align}
{Now, the normalization constant $\mathcal{N}_{j}$ becomes} {dependent on $j$.}

{Being similar} to the \cred{wave functions} on $T^2$, \cred{the results for the spinor are} applied to the scalar \cblue{case} as
\begin{gather}
\phi^j_{T^2/Z_2, n}(z) = \frac1{\sqrt 2} (\phi^j_{n} (z) + \eta (-1)^n \phi^{M-j}_{n}(z)).
\end{gather}
{Here, we immediately confirm the corresponding relations \cred{for} the mass eigenvalues\cred{,}}
\begin{gather}
\Delta \phi^j_{T^2/Z_2, n}(z) = m_n^2 \phi^j_{T^2/Z_2, n}(z),\\
m_n^2 = \frac{2\pi M}{\cal A} (2n+1).
\end{gather}
The multiplicity of the \cred{wave finctions} is also the same as that of {the} spinor.

\section{Brane-localized masses on \cred{a} magnetized orbifold background}
\label{sec:mass_perturbation}

\begin{figure}[t]
\centering
\includegraphics[width=0.7\textwidth]{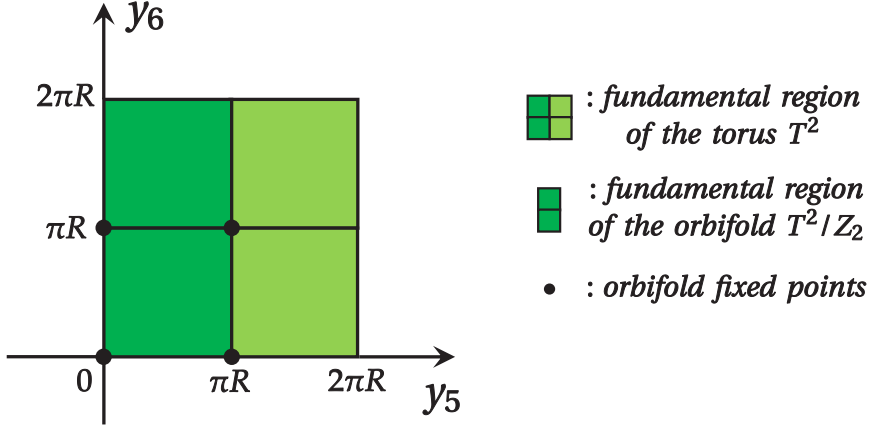}
\caption{The fundamental region of the orbifold $T^2/Z_2$ and {the} four fixed points {on it}.}
\label{fps}
\end{figure}

Before introducing brane-localized masses in magnetized extra dimensions, let us explain fixed points on toroidal orbifolds.
As explained already, toroidal orbifolds $T^2/Z_N$ for $N=2,3,4,6$ are obtained by an identification of {two-dimensional} extra dimensions under the toroidal periodicities and the $Z_N$ rotation,
\begin{gather}
z \sim z + 1 \sim z + {\tau} \sim e^{2\pi i/N} z,
\end{gather}
{where we keep the complex structure parameter in the general form.}
An important \cred{factor} in {extra-dimensional} model constructions is the presence of orbifold fixed points $z_{\rm fixed}$.
For $T^2/Z_2$, there exist four fixed points $z_{\rm fixed} = 0, 1/2, \tau/2, (1+\tau)/2$, as shown in Fig.\,\ref{fps}.
For the other orbifolds, the fixed points are {located} {at} {$z_{\rm fixed} = 0, (2+\tau)/3, (1+2\tau)/3$} for $T^2/Z_3$, {$z_{\rm fixed} = 0, (1+\tau)/2$} for $T^2/Z_4$ and $z_{\rm fixed} = 0$ for $T^2/Z_6$, respectively. 
Except for $T^2/Z_2$, {the} complex structure parameter should be discretized as $\tau = e^{2\pi i /N}$ due to consistency conditions of crystallography \cite{Choi:2006qh}.
For later convenience, the fixed points {of $T^2/Z_2$} are labeled as
\begin{gather}
z_1 = 0, \quad {z_2} = \frac12, \quad z_3 = \frac{i}{2}, \quad z_4 = \frac{1+i}2.
		\label{eq:T2Z2_fixedpoints}
\end{gather}
{We \cred{recall} that all of \cred{the} actual calculations are done \cred{with} the simple choice of $\tau = i$.}

\subsection{Scalar} 

We introduce a brane-localized mass \cred{at} a fixed point corresponding to {the} origin of two extra directions, i.e., $z = z_1 \, (= 0)$.
The Lagrangian for \cred{the} complex 6D scalar field under consideration is given as
\begin{gather}
{\cal L} = - | D_M \Phi (x^\mu, z) |^2 - h \, |\Phi(x^\mu, z)|^2 \delta^2({z - z_1}).
\label{action:scalar}
\end{gather}
Here, {the real dimensionless variable} $h$ denotes {the scalar mass localized} \cred{at} the fixed point, {where the mass scale is provided by the radius $R$}.
Also, this Lagrangian straightforwardly provides the {six-dimensional} equation of motion,
\begin{gather}
D_M D^M \Phi - h \, \Phi \delta^2({z - z_1}) = 0.
\end{gather}

By substituting the KK-expanded scalar \eqref{KKexpand2} {for} Eq.\,\eqref{action:scalar}, the effective Lagrangian after dimensional reduction is calculated as
\begin{align}
{\cal L}_{\rm eff} &= - \sum_{n, j} (\partial_\mu \varphi_n^j )^\dag (\partial^\mu \varphi_n^j) \notag\\
&\phantom{= \ } - \sum_{n, n'} \sum_{j, j'} (\varphi_n^j)^\dag \left( \int d^2z \, (D \phi^j_{T^2/Z_2, n}(z))^\dag \bigl(D \phi^{j'}_{T^2/Z_2, n'}(z) \bigr) + h (\phi^j_{T^2/Z_2, n}({z_1}))^\dag \phi^{j'}_{T^2/Z_2, n'}({z_1}) \right)\varphi_{n'}^{j'}
\notag\\
&\equiv {\cal L}_{\rm kin} - \sum_{n, n'} \sum_{j, j'} (\varphi_n^j)^\dag {\cal M}^2_{(n, j), (n',j')} \varphi_{n'}^{j'},
\end{align}
{where $\mathcal{L}_\text{kin}$ \cred{represents} the kinetic terms of KK scalar particles, which are canonically normalized.}
The explicit form of {the} KK mass matrix {after the perturbation} is given as
\begin{align}
{\cal M}^2_{(n, j), (n',j')} 
	&= {\frac{2\pi M}{\cal A} (2n+1) \delta_{n, n'} \delta_{j, j'} + h \,
	   (\phi^{j}_{T^2/Z_2, n}(z_1))^\dag (\phi^{j'}_{T^2/Z_2, n'}(z_1))} \notag \\
	&= {m_0^2} \, (2n+1) \delta_{n, n'} \delta_{j, j'} + 2h \, (v_{n, j})^\dag v_{n', j'} \quad
\label{massmat:scalar}
\end{align}
with {$m_0^2 = 2\pi M/{\cal A}$,} where we adopt \cred{the shorthand} notation
\begin{align}
v_{n, j} &\equiv \frac1{\sqrt2} \phi^j_{T^2/Z_2, n}({z_1}) = \phi^j_{n}({z_1}) \notag\\
&= \frac{{\cal N}_j}{\sqrt{{2^n n!}}} \sum_{\ell \in \mathbb{Z}} e^{-\pi M (\frac{j}{M} + \ell)^2} H_n \bigl( \sqrt{2\pi M} (\tfrac{j}{M} + \ell) \bigr).
\end{align}
{Here, we \cred{mention} the sign of the parameter $h$.
When $h$ is positive/negative, no/possible tachyonic modes appear in the spectrum.}

An extension to the localized {mass term} \cred{at} the other fixed {points} is straightforward; \cred{all we need to do is} change $z_1$ \cred{to} $z_i \,\, (i=2,3,4)$ in $v_{n,j}$.
Here, we comment on the fourth fixed point $z_4$ and effects from {the} localized mass on it.
Via direct {calculations}, we obtain
\begin{align}
\phi^j_{T^2/Z_2, n}(-z_k) = (-1)^{M \delta_{k, 4}} \, {\eta} \, \phi^j_{T^2/Z_2, n} (z_k)
\end{align}
for $k=1,2,3, 4$.
This implies that when $\eta=-1$ there is no effect\cred{,} except for $k=4$ with odd $M$.\footnote{
In \cred{other words}, when $\eta = +1$, a brane-localized mass term manifestly vanishes only in the case of $k=4$ with odd $M$ (see also \cred{Table} \ref{defofalphaM3}).
}
In this paper, we restrict ourselves to {the} case of $\eta = +1$.
Such brane-localized mass terms of the scalar field possibly {affect} physics at \cred{low energies}.
Hence, we are interested in {the eigenvalues and eigenvectors of the perturbed mass matrix in Eq.\,(\ref{massmat:scalar})}.
{We note that the above relation is also derived from the \cred{pseudoperiodic} boundary conditions on \cred{both} $T^2$ \cred{and} $T^2/Z_2$ (see\cred{,} e.g., \cred{Ref.}\ \cite{Abe:2013bca})\cred{,}
\begin{align}
& \phi^j_{T^2  \text{ or } T^2/Z_2, n}(z+1) = e^{i q \chi_1(z)} \phi^j_{T^2 \text{ or } T^2/Z_2, n}(z), \\
& \phi^j_{T^2  \text{ or } T^2/Z_2, n}(z+i) = e^{i q \chi_i(z)} \phi^j_{T^2 \text{ or } T^2/Z_2, n}(z), \\
& \chi_1(z) =  \frac{b}{2} \text{Im}(z),\quad
  \chi_i(z) = -\frac{b}{2} \text{Im}(i z),
\end{align}
by setting $z$ as $0$, $-1/2$, $-i/2$, \cred{or} $-(1+i)/2$.}
\cred{We will} diagonalize this KK mass matrix \cred{in the following}.

First of all, we comment on \cred{the} cutoff scale.
Unfortunately, {extra-dimensional} models are \cred{nonrenormalizable}, and hence they should be considered as {a kind} of LEET of {a} more fundamental theory, e.g., string theory at \cred{a} scale {below} \cred{the} cutoff scale $\Lambda$.
In {the following discussion}, we define the cutoff scale as a {certain} level of the KK masses, i.e.,
\begin{gather}
m_0^2 < m_1^2 \, (= 3m_0^2) < ... < {m^2_{n_{\text{max}}}} \equiv \Lambda^2 < {m^2_{n_{\text{max}}+1}} < {\cdots},
\end{gather}
where $\Lambda$ is related to the size of the {perturbed} KK mass matrix by turning on the brane-localized mass term.

\cred{Also, a few comments about} the cutoff of theories in extra dimensions \cred{are in order}.
When the reference energy in \cred{the} renormalization group evolution crosses \cred{the} mass of KK particles, beta functions take contributions from the states.
Cumulative spectra of KK particles, which {are} a generic structure of compact extra dimensions, {lead} to \cred{a rapid increase and decrease} of effective 4D couplings immediately once the reference scale passes the lowest KK state (see\cred{,} e.g., \cred{Ref.}\ \cite{Dienes:1998vh,Dienes:1998vg,Kakuda:2013kba}).
\cred{Therefore, in general} the cutoff scale should be close to \cred{the} typical size of KK particles.
When \cred{the} cutoff scale is not \cred{very} far away from the electroweak scale, corrections via higher-dimensional operators are not suppressed.
On the other hand, when \cred{the} KK mass scale is far away from the electroweak scale, such contributions are subdued (though the cutoff scale is close to \cred{the} typical scale of KK states).
Hence, {for} an extra-dimensional theory with \cred{a} sufficiently heavier KK mass scale compared with the scale of \cred{electroweak} physics, the fact that the cutoff scale should be near \cred{the} typical scale of KK particles \cred{does not seem to be} problematic.

It is convenient to relabel {the indices} {$(n^{(\prime)},j^{(\prime)})$} in the KK mass matrix {in Eq.\,\eqref{massmat:scalar}} in terms of a new label $\alpha$.
We {describe} the degeneracy of {the} \cred{wave functions} for even $n$ (odd $n$) {by} $d_{e}$ ($d_{o}$), {\cred{which gives us} the useful \cred{relation}} $M = d_e + d_o$.
{We} can define {a} one-to-one labeling as {shown in} \cred{Table} \ref{defofalpha}.\footnote{
{When $\eta = +1$ and $n$ is odd, mode functions vanish in $j=0$.
\cred{Therefore}, $j$ starts from one (not zero) in the category $n=1$ in \cred{Table} \ref{defofalpha}.}
}
For example, when $M=3$, {an explicit correspondence} between $\alpha$ and $(n, j)$ ({up to the} {ninth} mode) is shown in \cred{Table} \ref{defofalphaM3}.
{In terms of} this labeling, the KK mass matrix (``wavefunction vector'') is expressed as ${\cal M}^2_{(n,j), (n',j')} \to {\cal M}^2_{\alpha, \beta}$ ($v_{n,j} \to v_\alpha$), respectively. 
{Also, by use of the information \cred{in Tables} \ref{numofneven} and \ref{numofnodd}, the size of the mass matrix is easily estimated as}
\begin{align}
{
\alpha_{\text{max}} =
\begin{cases}
M \times (k-1) + d_e & \text{for } n_{\text{max}} = 2(k-1), \\
M \times  k          & \text{for } n_{\text{max}} = 2k-1,
\end{cases}
\qquad
\left( k = 1,2,3,\cdots \right),}
			\label{eq:alpha_max}
\end{align}

\begin{table}[t]
\centering
\vspace{15pt}
\begin{tabular}{|c|cccc|cccc|cccc|c|} \hline
$\alpha$ & $1$ & $2$ & ... & $d_e$ & $d_e+1$ & $ d_e+2$ & ... & $M$ & $M+1$ & $M+2$ & ... & $M + d_e$ &...\\ \hline
$n$ & $0$ & {$0$} & ... & $0$ & $1$ & $1$ & ... & $1$ & $2$ & $2$ & ... & $2$ & ...\\
$j$ & $0$ & $1$ & ... & $d_e -1$ & $1$ & $2$ & ... & $d_o$ & $0$ & $1$ & ... & $d_e-1$ & ...\\ \hline
\end{tabular}
\caption{Relationships between $\alpha$ and $(n,j)$.}
\label{defofalpha}
\vspace{15pt}
\begin{tabular}{|c|cccccccccc|} \hline
$\alpha$ & $1$ & $2$ & $3$ & $4$ & $5$ & $6$ & $7$ & $8$ & $9$ & ... \\ \hline
$(n,j)$ & $(0,0)$ & $(0,1)$ & $(1, 1)$ & $(2,0)$ & $(2,1)$ & $(3,1)$ & $(4,0)$ & $(4,1)$ & $(5,1)$ & ... \\ \hline
\end{tabular}
\caption{Relationships between $\alpha$ and $(n,j)$ for $M=3$.}
\label{defofalphaM3}
\end{table}

Now, we are ready to analyze the {perturbed} KK mass matrix.
{As \cred{a} first} illustration, let us consider $M=3$ and $\Lambda = m_1$, {where $d_e=2$, ${d_o}=1$, and $\alpha$ runs over $1, 2, 3$}.
{The corresponding} $3 \times 3$ KK mass matrix is given as
\begin{gather}
{\cal M}^2_{\alpha, \beta} =
\begin{pmatrix}
m_0^2 + 2 h |v_1|^2 & 2h v_1^\dag v_2 & 2h v_1^\dag v_3\\
2h v_2^\dag v_1 & m_0^2 + 2 h |v_2|^2 & 2h {v_2^\dag v_3}\\
2h v_3^\dag v_1 & 2h v_3^\dag v_2 & m_1^2 + 2h |v_3|^2
\end{pmatrix},
\label{3x3scalar}
\end{gather}
and three eigenvalues can be analytically solved as
\begin{gather}
m_0^2, \quad 2 m_0^2 + h(|v_1|^2 + |v_2|^2 + |v_3|^2) \pm \sqrt{(m_0^2 + h (|v_1|^2 + |v_2|^2 + |v_3|^2))^2 -4h m_0^2 (|v_1|^2 + |v_2|^2)},
\end{gather}
{where we \cred{used} the relation $m_1^2 = 3 m_0^2$.}
{Here, \cred{the} eigenvalue is unperturbed as $m_0^2$.}
By {focusing the property $|v_\alpha R|^2 \lesssim 1$}, {which is recognized by the correct normalization of the mode functions,} the other eigenvalues are roughly estimated as {$m_0^2(1 + {\cal O}(h))$ and $m_1^2(1 + {\cal O}(h))$}.
Thus, we find that one of the original lowest KK {masses}, i.e., $m_0^2$ appears after turning on the brane-localized mass.

When the cutoff scale is chosen as $\Lambda = m_3$, \cred{where} $\alpha=1,2, ..., 6$, \cred{the} corresponding eigenvalues are calculated {in a similar manner} as {$m_0^2$, $m_0^2(1 + {\cal O}(h))$, $m_1^2(1 + {\cal O}(h))$, $m_2^2$, $m_2^2(1 + {\cal O}(h))$\cred{,} and $m_3^2(1 + {\cal O}(h))$}.
\cred{We find that} {one of the second excited} \cblue{states} is also unperturbed.

The above discussion can be extended to {the} generic {magnitude} {of} the magnetic flux and an arbitrary cutoff scale.
The KK mass matrix with a brane-localized mass can be symbolically expressed as
\begin{gather}
{\cal M}^2 = {{\rm diag} \, (\underbrace{m_0^2, m_0^2, ..., m_0^2}_{d_e}, \, \underbrace{m_1^2,m_1^2,...,m_1^2}_{d_o}, \, \underbrace{m_2^2, m_2^2, ..., m_2^2}_{d_e}, \, \cdots )} + v^\dag \otimes v,
\label{symb}
\end{gather}
where ${(v)_\alpha} = (v_1, v_2, ...)$ denotes {an} {$\alpha_\text{max}$} component complex vector.
For $d_e \geq 2$, there always exists an eigenvector {for the lowest mode ($n=0$)},
\begin{gather}
(u_{{n=0}})_\alpha = (u_1, u_2,..., u_{d_e}, 0, 0, \cdots),
\end{gather}
{which \cred{satisfies}} $v \cdot u_{{n=0}} =0$.
In the above example ($M=3$), {there exist $d_e-1$ linearly independent eigenvectors} that satisfy $v \cdot u_{{n=0}} =0$ as
\begin{align}
\left(u^{{(1)}}_{{n=0}}\right)_\alpha &= (\underbrace{-v_2, v_1, 0, 0, ..., 0}_{d_e}, \, 0, 0, \cdots), \label{eigen1}\\
\left(u^{{(2)}}_{{n=0}}\right)_\alpha &= (\underbrace{0, -v_3, v_2, 0, ..., 0}_{d_e}, \, 0, 0, \cdots),\\
\vdots\notag\\
\left(u^{{(d_e -1)}}_{{n=0}}\right)_\alpha &= (\underbrace{0,0, ..., 0, -v_{d_e}, v_{d_e-1}}_{d_e}, \, 0, 0, \cdots), \label{eigende-1}
\end{align}
where we cannot take another eigenvector \cred{that is} linearly independent of all of {$u^{(1)}_{{n=0}}$,\,$u^{(2)}_{{n=0}}$,\,...,\,$u^{(d_e -1)}_{{n=0}}$}.
This fact suggests that one of the lowest modes is uplifted by the perturbation {after {turning} on the brane-localized mass}.
For any level of the {degenerate} mass eigenvalues {before the perturbation}, we find that such vectors provide the {corresponding} eigenvalue as
\begin{gather}
{{\cal M}^2 u_{n} = m_n^2 u_n + v^\dag (v \cdot u_n) = m^2_n u_n}.
\end{gather}

We would like to comment on the effects from multiple localized {mass terms}.
For example, we turn on two localized masses $h_1|\Phi|^2$ and $h_2|\Phi|^2$ at $z=z_1$ and $z=z_2$, respectively.
{Here,} the corresponding KK mass matrix \cred{is}
\begin{gather}
\left({\cal M}^2\right)_{\alpha, \beta} = (2n+1)m_0^2 \delta_{\alpha, \beta} + 2h_1 v^{(1)\dag}_{\alpha} v^{(1)}_{\beta} + 2h_2 v^{(2)\dag}_{\alpha} v^{(2)}_{\beta},
\end{gather}
where we define
\begin{align}
v^{(1)}_{\alpha} &\equiv v^{(1)}_{n,j} = \phi^j_{n}(z_1),\\
v^{(2)}_{\alpha} &\equiv v^{(2)}_{n,j} = \phi^j_{n}(z_2).
\end{align}
{The} \cred{degeneracies} of the KK {states} via magnetic fluxes \cred{are} {degraded} one by one as we place the brane-localized mass.
{We provide a detailed discussion on such cases in \cred{Appendix} \ref{app:multiple_massterm}.}

Before closing this section, it is important to mention {a 6D vector field, which is decomposed into a (4D) vector component ($A_{\mu}$) and two scalar components ($A_{5,\,6}$) {from} the four-dimensional point of view}.
The KK mass spectra of a vector field {which feels magnetic fluxes} on \cred{a} flux background \cred{were} analyzed in Ref.\,\cite{Hamada:2012wj}, which \cred{provided} KK eigenvalues {of the two corresponding 4D scalars of} $\phi_{n, z} \equiv (A_5 + i A_6)/\sqrt2$ {as}
\begin{gather}
m_n^2 = \frac{2\pi M}{\cal A} (2n-1),
	\label{eq:Az_spectrum}
\end{gather}
{and of} $\phi_{n, \bar z} \equiv (A_5 - i A_6)/\sqrt2$ {as}
\begin{gather}
m_n^2 = \frac{2\pi M}{\cal A} (2 (n+1) +1),
\end{gather}
{where the spectrum of the vector field is equivalent to $\phi_{n, \bar z}$ through suitable gauge fixing.
Here, we would pay attention to two points.
One is that the equation of motion of the 6D scalar $\Phi$ {takes} \cred{a different form than} those of $\phi_{n, z}$ and $\phi_{n, \bar z}$, which leads to the difference in the mass spectra.
The other is that in the present Abelian case, the 6D vector field does not feel {any} magnetic flux.
The situation \cred{where} {the} 6D vector field couples to magnetic flux is realized in {a non-Abelian gauge theory}, which is a reasonable playground for unified theories.}

{From Eq.~(\ref{eq:Az_spectrum}), we \cred{recognize} a pathology of the emergence of \cred{the} tachyonic state in $\phi_{n=0, z}$, which is a critical obstacle for constructing reasonable models.
An ordinary remedy for {conquering} the difficulty is to address \cred{supersymmetrized} theories, where if 4D $\mathcal{N} = 1$ supersymmetry remains in {the} action (before taking into account \cred{the connection to the supersymmetry-breaking} sector), such tachyonic states are stabilized (see\cred{,} e.g., \cred{Refs.} \cite{Cremades:2004wa,Hamada:2012wj}).
On the other hand, issues discussed \cred{in} this manuscript would provide another clue to \cred{circumventing the obstacle by} {uplifting} the tachyonic mode via a brane-localized mass term for $\phi_{n, z}$.}

When localized mass terms, e.g., $h A_5^2$ ({and} $h A_6^2$) {are} induced after {non-Abelian gauge symmetry breaking} \cred{(}i.e., introducing fluxes and/or Wilson lines\cred{)}\footnote{
Actually in the present situation, an Abelian gauge field (and Cartan parts of a non-Abelian gauge field) cannot feel magnetic fluxes and \cred{thus} the lowest mode remains massless, while non-Cartan parts of a non-Abelian gauge field can detect such fluxes and \cred{can} be massive.
Also, we mention that {a} gauge-invariant mass term via {(classical) flux configurations} cannot be confined within a 4D world, which means that other extra spacial direction(s) should be required in addition to the present two directions to realize the localized mass term of $\phi_{n,z}$ and $\phi_{n,\bar{z}}$ in a \cred{gauge-invariant} way.} 
{the} {above discussion is} relevant \cred{for analyzing} the KK mass matrix of such {kinds} of scalars, in spite of the difference in the pattern of the KK masses.

\subsection{Spinor}
\label{sec:spinor}

We express the 6D Weyl spinor as $\Psi = (\lambda_+, \lambda_-)^\text{T}$ in terms of \cred{four-component} Weyl spinors $\lambda_+$ and $\lambda_-$\cred{. Then,} its KK decomposition is given as
\begin{align}
\lambda_+ (x^\mu, z) = {\sum_{n,j} \chi_{+, n}^j (x^\mu) \otimes \psi_{T^2/Z_2 \, +, n}^j(z)}, \label{kkspinor1}\\
\lambda_- (x^\mu, z) = {\sum_{n,j} \chi_{-, n}^j (x^\mu) \otimes \psi_{T^2/Z_2 \, -, n}^j(z)}. \label{kkspinor2}
\end{align}
The Lagrangian for {the} 6D Weyl spinor is given as
\begin{gather}
{\cal L} = i \bar \Psi \Gamma^M D_M \Psi + (i \bar \nu \gamma^\mu \partial_\mu \nu -g (\bar \nu \lambda_+ + {\rm \cred{H}.c.})) \delta^2(z-z_k),
\label{action:spinor}
\end{gather}
where $g$ is a massless parameter associated with \cred{the} localized mass of the spinor.
Note that the {6D} Weyl spinor cannot possess a Dirac mass term such as $\bar \Psi \Psi$ \cite{Dudas:2005vn}.
Then, we add a four-dimensionally localized {Weyl} spinor field $\nu(x^\mu)$ and introduce a localized mass term \cred{at} a fixed point $z_k$.

In the following\cred{, we place the localized mass at} {a fixed point of $T^2/Z_2$ [see Eq.\,(\ref{eq:T2Z2_fixedpoints})]}.
It is easily found that $\lambda_- (x^\mu, {z_k})=0$ ($\lambda_+(x^\mu, {z_k})=0$) {[${k}=1,2,3$]} for $\eta=+1$ ($\eta=-1$), respectively.
Hence, we {cannot} introduce such a mass for $\lambda_-$ {\cred{at} the three fixed points}.
\cred{Therefore}, we focus on the case of $\eta=+1$, \cred{as} in the previous section.
It is straightforward to expect that we can similarly analyze \cred{the} case of $\eta=-1$.
Hereafter, we choose the 6D chirality of $\Psi$ as $-1$, which means that left-handed chiral modes are realized as zero modes of $\lambda_+$ (see \cred{Appendix} \ref{app:notation}).
The 4D chirality of $\nu$ is automatically determined \cred{to be} {right-handed}.

The spinor Lagrangian in Eq.\,\eqref{action:spinor} provides the {six-dimensional} equations of motion,
\begin{align}
i \gamma^\mu \partial_\mu \lambda_+ + D^\dag \lambda_- - g \nu \delta^2({z - z_k}) &=0,\\
i \gamma^\mu \partial_\mu \lambda_- - D \lambda_+ &= 0,\\
{\left( i \gamma^\mu \partial_\mu \nu - g {\lambda_{+}} \right)} \delta^2(z - z_k) &=0.
\end{align}
{Removing $\lambda_-$ and $\nu$} {from} these equations leads to the {(six-dimensional)} equation of motion for $\lambda_+$\cred{,}
\begin{gather}
{\left( \partial_\mu \partial^\mu + D^\dagger D \right)} \lambda_+ - {g^2} \lambda_+ \delta^2({z - z_k}) =0,
\end{gather}
{where the operator $D^\dagger D$ is equal to $\Delta - 2 \pi M/\mathcal{A}$, which indicates the mass difference in the fermion case and {scalar} case as shown in Eqs.~(\ref{eq:KKmass_fermion}) and (\ref{eq:KKmass_scalar}) when $g=0$.}

By plugging the KK decompositions \eqref{kkspinor1} and \eqref{kkspinor2} into Eq.\,\eqref{action:spinor}, the effective Lagrangian under \cred{the} cutoff scale is calculated as
\begin{align}
{\cal L}_{\rm eff} 
&= {\cal L}_{\rm kin} - \sum_{{n, j \atop (n \not= 0)}} \left(m_n {\bar \chi}^j_{+, n}{}_L \chi^{j}_{-, n}{}_R  + g {\bar \nu}_R \chi_{+, 0}^j{}_L \psi^j_{T^2/Z_2 \, +, 0}({z_k}) + g {\bar \nu}_R \chi^j_{+,n}{}_L \psi^j_{T^2/Z_2 \, +, n}({z_k}) + {\rm \cred{H}.c.} \right) \notag\\
&\equiv {\cal L}_{\rm kin} - \sum_{{n, j \atop (n \not= 0)}} (\bar \nu_R, \, {\bar \chi}_{-,n}^j{}_R) {\cal M}
\begin{pmatrix}
{\chi}^j_{+,0}{}_L\\[3pt]
{\chi}^j_{+,n}{}_L
\end{pmatrix}
+ {\rm \cred{H}.c.},
\end{align}
where {$\mathcal{L}_\text{kin}$ contains kinetic terms and} the corresponding 4D chiralities are explicitly shown for clarity.
The perturbed KK mass matrix $\mathcal{M}$ under the brane-localized spinor mass term is symbolically expressed as
\begin{gather}
{\cal M} = 
\begin{pmatrix}
\sqrt2 \, {g} \psi^j_{+,0}({z_k}) & \sqrt2 \, {g} \psi^j_{+,n}({z_k}) \\[3pt]
0 & m_n
\end{pmatrix}.
\end{gather}
Here, note that \cred{the} index $n$ takes nonzero {positive} \cred{integer values} in the above expressions\cred{,} and also that we use \cred{the} relation $\psi^j_{T^2/Z_2 \, +, n}({z_k}) = \sqrt2 \psi^j_{+,0}({z_k})$.
{Here, the size of $\mathcal{M}$ is $(1 + N_\text{KK}) \times (d_e + N_\text{KK})$, where $N_\text{KK}$ represents {the number of excited KK modes} \cred{that} appear up to the level designated by $n_\text{max}$.
The relation is easily understood among $N_\text{KK}$ and $\alpha_\text{max}$ {as} defined in Eq.\,(\ref{eq:alpha_max}),}
\begin{align}
{N_\text{KK} = \alpha_\text{max} - d_e,}
\end{align}
because $d_e$ {represents} the number of \cred{zero modes}.

Since the matrix $\mathcal{M}$ is asymmetric, it is convenient to consider the following two forms of products of {the matrix}
\begin{align}
({\cal MM}^\dag)_{\alpha, \beta} 
	&= {(\Pi^2_R)_{\alpha,\beta} +
	   \sqrt{2} \, {g} (V^\dag_R \Pi_R)_\alpha \, \delta_{0, \beta} + 
	   \sqrt{2} \, {g} (\Pi_R V_R)_\beta \, \delta_{0, \alpha}} \notag \\
	&\quad
	   {+ 2 {g^2}
	   \left[
	   \sum_{j=0}^{d_e -1} |v_{0,j}|^2 +
	   \sum_{\rho = 1}^{N_\text{KK}} |(V_R)_\rho|^2
	   \right] {\delta_{0, \alpha} \delta_{0, \beta}}
	   \qquad (\alpha, \beta = 0,1,\cdots,N_\text{KK})},
\label{MMdag} \\
({\cal M}^\dag {\cal M})_{\alpha, \beta} 
	&= {(\Pi^2_L)_{\alpha, \beta} + 
	   \sqrt{2} \, {g} (V^\dag_R)_\alpha (V_R)_\beta \qquad (\alpha, \beta = 1,2,\cdots,\alpha_\text{max} {=N_\text{KK}+d_e})},
\label{MdagM}
\end{align}
for {the right- and left-handed Weyl spinors} $(\nu_R, {\chi}_{-,n}^j{}_R)$ and $({\chi}^j_{+,0}{}_L, {\chi}^j_{+,n}{}_L)$, respectively.
Here, we \cred{see} that $\alpha=0$ corresponds to the 4D localized field {$\nu_R$} in Eq.\,\eqref{MMdag}.
{The sizes of the matrices ${\cal MM}^\dag$ and ${\cal M}^\dag {\cal M}$ are $(1 + N_\text{KK})\times(1 + N_\text{KK})$ and $(d_e + N_\text{KK})\times(d_e + N_\text{KK})$, respectively.}
{Also,} we define the following symbols
\begin{align}
(\Pi_R)_{\alpha, \beta} 
	&\equiv {\rm diag} 
	        (0, \, 
	               \underbrace{m_1, m_1, ..., m_1}_{d_o}, \, 
	               \underbrace{m_2, m_2, ..., m_2}_{d_e}, \cdots), \\
(V_R)_\alpha 
	&\equiv (0, \, 
	               \underbrace{v_{1, 1}, v_{1, 2}, ..., v_{1, d_o}}_{d_o}, \, 
	               \underbrace{v_{2, 0}, v_{2, 1}, ..., v_{2, d_e-1}}_{d_e}, \, \cdots)^\text{T}, \\
(\Pi_L)_{\alpha, \beta} 
	&\equiv {\rm diag} 
	        (      \underbrace{0, 0, ..., 0}_{d_e}, \, 
	               \underbrace{m_1, m_1, ..., m_1}_{d_o}, \, 
	               \underbrace{m_2, m_2, ..., m_2}_{d_e}, \cdots), \\
(V_L)_\alpha 
	&\equiv (      \underbrace{v_{0, 0}, v_{0, 1}, ..., v_{0, d_e-1}}_{d_e}, \, 
	               \underbrace{v_{1, 1}, v_{1, 2}, ..., v_{1, d_o}}_{d_o}, \, 
	               \underbrace{v_{2, 0}, v_{2, 1}, ..., v_{2, d_e-1}}_{d_e}, \, \cdots)^\text{T},
\end{align}
with $v_{n, j} \equiv \psi^j_{+, n}({z_k})$.

In the spinor case, we recognize that the mass spectrum of the left-handed modes $\chi_{+, 0}^j{}_L$ and $\chi_{+, n}^j{}_L$  is equivalent to that of the scalar {because the mass matrix squared in Eq.\,(\ref{MdagM}) takes the same form as that in the scalar case}.

On the other hand, we analyze \cred{the} mass spectra of the 4D brane-localized field $\nu_R$ and the right-handed modes ${\chi}_{-,n}^j{}_R$.
For a vector $(u)_\alpha \,\, (\alpha = 0,1,\cdots,N_\text{KK})$, we calculate \cred{the} product of $({\cal MM}^\dag)_{\alpha, \beta}$ and $u_\alpha$ as
\begin{gather}
\sum_{\beta=0}^{N_\text{KK}} ({\cal MM}^\dag)_{\alpha, \beta} (u)_\beta 
=
\begin{cases}
\displaystyle \sqrt{2} \, {g} \sum_{\beta=1}^{N_\text{KK}} (\Pi_R V_R)_\beta (u)_\beta + 
		2 {g^2} \left[ \sum_{j=0}^{d_e -1} |v_{0,j}|^2 + 
		            \sum_{{\rho}=1}^{N_\text{KK}} |(V_R)_\rho|^2 \right] (u)_0   &   (\alpha=0),\\[5pt]
\displaystyle \sum_{\beta=0}^{N_\text{KK}} (\Pi^2_R)_{\alpha, \beta} (u)_\beta + 
		\sqrt{2} \, {g} (V^\dag_R \Pi_R)_\alpha (u)_0  &  (\alpha \geq 1).
\end{cases}
\label{spinormass}
\end{gather}
\cred{Equation} \eqref{spinormass} implies that {$\nu_R$} is perturbed by the presence of the localized mass since the {right-hand} side for $\alpha=0$ {may be} \cred{nonvanishing} in almost all cases.
This is understood as follows.
Since the right-handed spinors $\chi^j_{-, n}{}_R$ are originally massive around the compactification scale ($\sim 1/\sqrt{\cal A}$), we can \cred{determine} whether {$\nu_R$} is massless or massive only by investigating {the} determinant of ${\cal MM}^\dag$.
This is because {the} determinant of a matrix \cred{is equal} to {the} product of its eigenvalues.
If the determinant is {nonzero}, we can conclude that {$\nu_R$} {becomes} massive.

{Here}, let us focus on {a} {simple} example for $M=3$ and $\Lambda=m_2$.
For {the right-handed fields} $(\nu_R, {\chi}_{-,1}^{0}{}_R, {\chi}_{-,1}^{1}{}_R, {\chi}_{-,2}^{1}{}_R)$, the $4 \times 4$ KK mass matrix {squared} ${\cal MM}^\dag$ is symbolically written \cred{with the} symbols $a$, $b$, $c$ and $d$ as 
\begin{gather}
{\cal MM}^\dag =
\begin{pmatrix}
{2} \, {g^2} d & {\sqrt{2}} \, {g} a \,{m_1} & {\sqrt{2}} \, {g} b \,{m_1} & {\sqrt{2}} \, {g} c \,{m_2} \\
{\sqrt{2}} \, {g} a^\dag {m_1} & {m_1^2} & 0& 0\\
{\sqrt{2}} \, {g} b^\dag {m_1} & 0 & {m_1^2} & 0\\
{\sqrt{2}} \, {g} c^\dag {m_2} & 0 & 0 & {m_2^2}
\end{pmatrix}.
\end{gather}
Here, $g$ is a dimensionless coefficient in the localized mass term of the spinor field.
The determinant of this matrix is calculated as
\begin{gather}
{\det\left({\cal MM}^\dag\right) = 2 {g^2} m_1^4 m_2^2 
	\left(
	d - |a|^2 - |b|^2 - |c|^2
	\right),}
\end{gather}
where mass dimensions of the four symbols are two (for $d$) and one (for {$a, b, c$}).
For a {nonzero} {coefficient ($g \neq 0$), a miraculous cancellation should \cred{occur} among the symbols $a, b, c$ and $d$ to realize $\det {\cal MM}^\dag =0$, which suggests that {$\nu_R$} is still massless after the {localized-mass} perturbation}.
{Although we} note that we could not find a concrete example where the cancellation happens{, our conclusion is} that {$\nu_R$} gets a mass via the localized mass in almost any case.
When $M$ and/or $\Lambda$ is {arbitrary}, configurations of nonzero components in the mass matrix squared $\mathcal{M} \mathcal{M}^\dagger$ {look} similar.
\cred{Therefore}, this kind of discussion is still valid.

{The matrices ${\cal MM}^\dag$ and $\mathcal{M}^\dagger \mathcal{M}$ contain $N_{\text{KK}} + 1$ and $N_{\text{KK}} + d_e$ numbers of squared mass eigenvalues, respectively, where $N_\text{KK}$ values are common in both of the matrices. These modes mainly originate from the KK mass terms \cred{$\overline{\chi}^j_{-,nR} \chi^j_{+,nL} + \text{H.c.}$} which exist {even} before the perturbation.}

\section{Deformations of the KK \cred{wave functions}}
\label{sec:wavefunction}

As discussed in Sec.\,\cblue{3}, we {found} that \cred{at least} one of the lowest KK masses \cred{remains} after introducing {one brane-localized mass term if the number of the lowest modes is \cred{greater} than or equal to two}.
In this section, we investigate what happens on the corresponding KK \cred{wave function} {after the perturbation}.

A physical sense of such deformations is a possible modulation of three-point effective interactions, i.e., Yukawa couplings in {phenomenological models, which are} characterized by the overlap integrals of three types of KK wavefunctions.\footnote{For example, see Refs.\,\cite{Cremades:2004wa, Abe:2008sx} and also Ref.\,\cite{Ibanez:2012zz}.}
The Yukawa couplings on $T^2/Z_2$ are expressed as
\begin{gather}
y_{{\alpha\beta\gamma}} \sim \int_{{T^2}} d^2z \, \psi_{T^2/Z_2}^{{\alpha}}(z) \psi_{T^2/Z_2}^{{\beta}} (z) (\phi_{T^2/Z_2}^{{\gamma}}(z))^\dag,
\end{gather}
{where $\alpha$, $\beta$ and $\gamma$ discriminate the physical eigenstates on magnetized $T^2/Z_2$ before the perturbation by} \cred{turning on} single or multiple \cred{brane-localized} mass terms.
For example, when we introduce a localized mass {term for} the 6D scalar field, the profiles of {KK mode functions describing the lowest states} would be changed as $\phi_{T^2/Z_2}^{\alpha}(z) \,\, ({\alpha} = {1,2, ..., {d_e}}) \to \phi^{{i_\text{mass}}}_{{T^2/Z_2}}(z) \,\, ({i_\text{mass}} ={1,2, ..., {d_e} -1})$.
Accordingly, the Yukawa couplings are expected to be changed as 
\begin{gather}
\int_{{T^2}} d^2z \, \psi_{T^2/Z_2}^{{\alpha}}(z) 
                          \psi_{T^2/Z_2}^{{\beta}} (z) 
                         (\phi_{T^2/Z_2}^{{\gamma}}(z))^\dag 
	\to 
\int_{{T^2}} d^2z \, \psi_{T^2/Z_2}^{{\alpha}}(z) 
                          \psi_{T^2/Z_2}^{{\beta}} (z) 
                         (\phi_{{T^2/Z_2}}^{{i_\text{mass}}} (z))^\dag.
\end{gather}
It is also expected that the same holds for localized mass terms of the spinor.
Although we do not analyze such Yukawa couplings in the presence of brane-localized masses in this paper, it is important to investigate the change of the lowest KK mode functions.

\begin{figure}[H]
\centering
\includegraphics[width=0.3\textwidth]{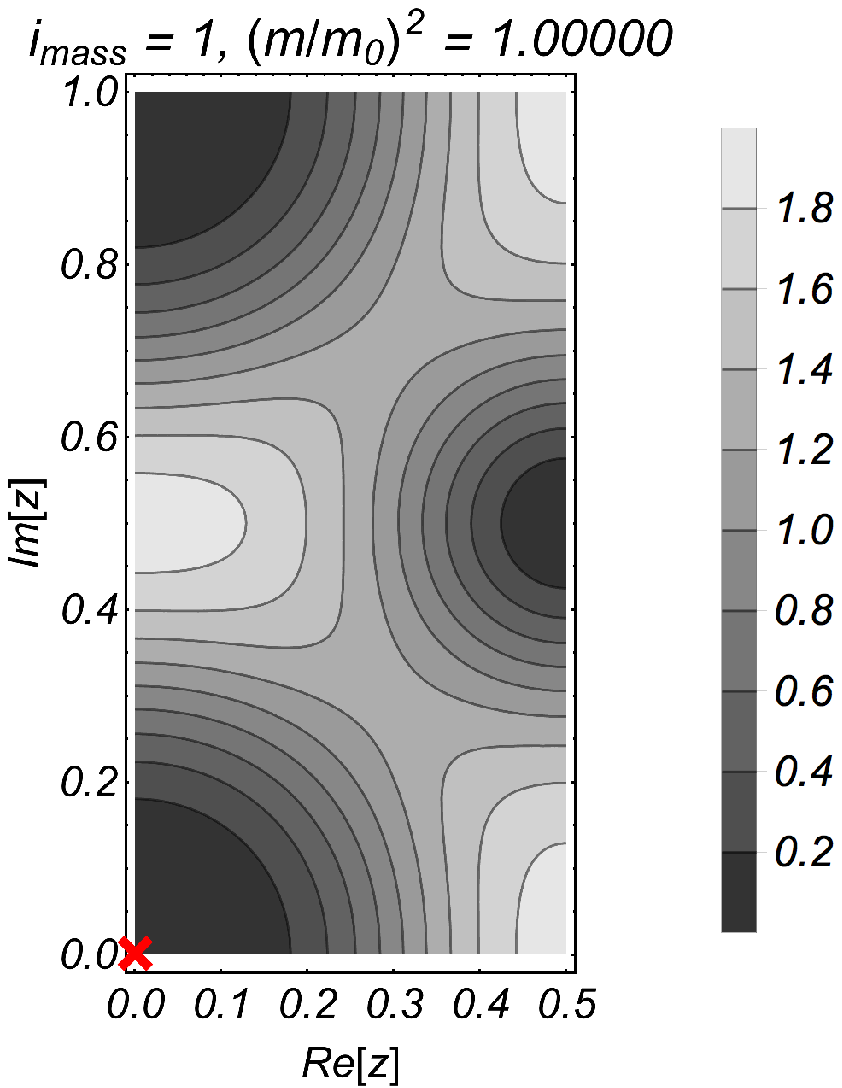}\hspace{0.033\textwidth}
\includegraphics[width=0.3\textwidth]{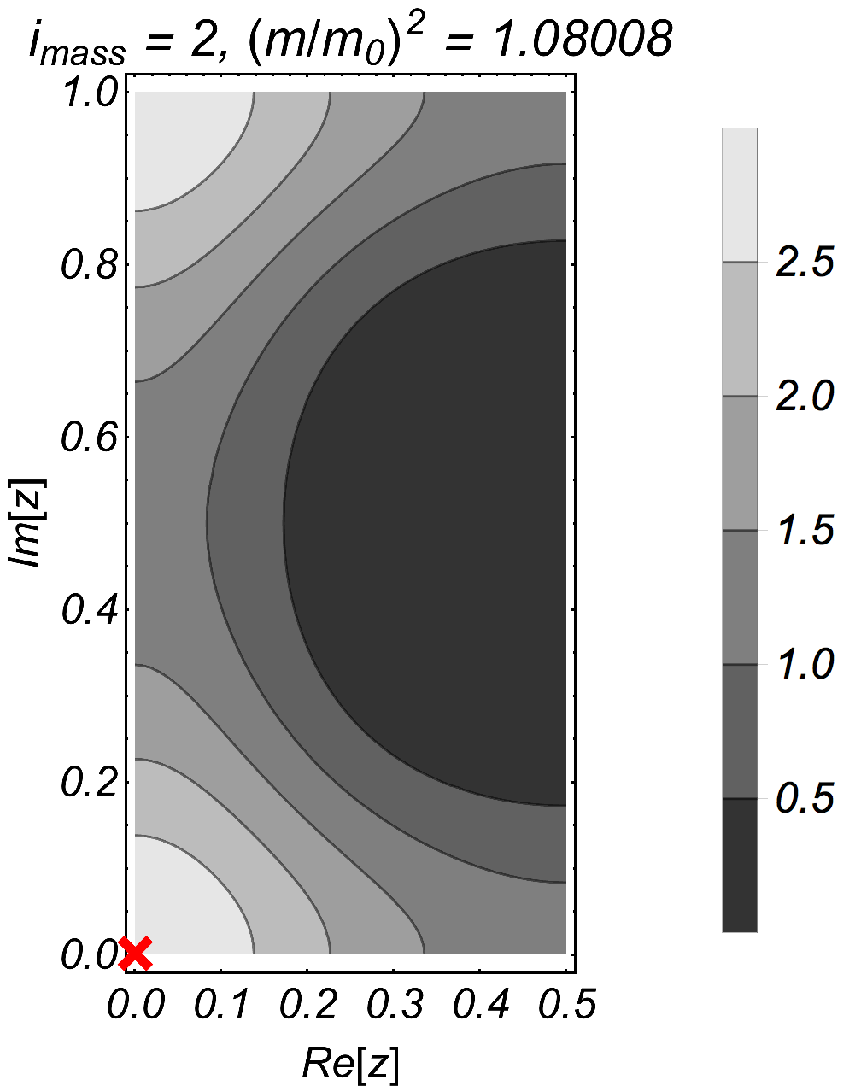}\hspace{0.033\textwidth}
\includegraphics[width=0.3\textwidth]{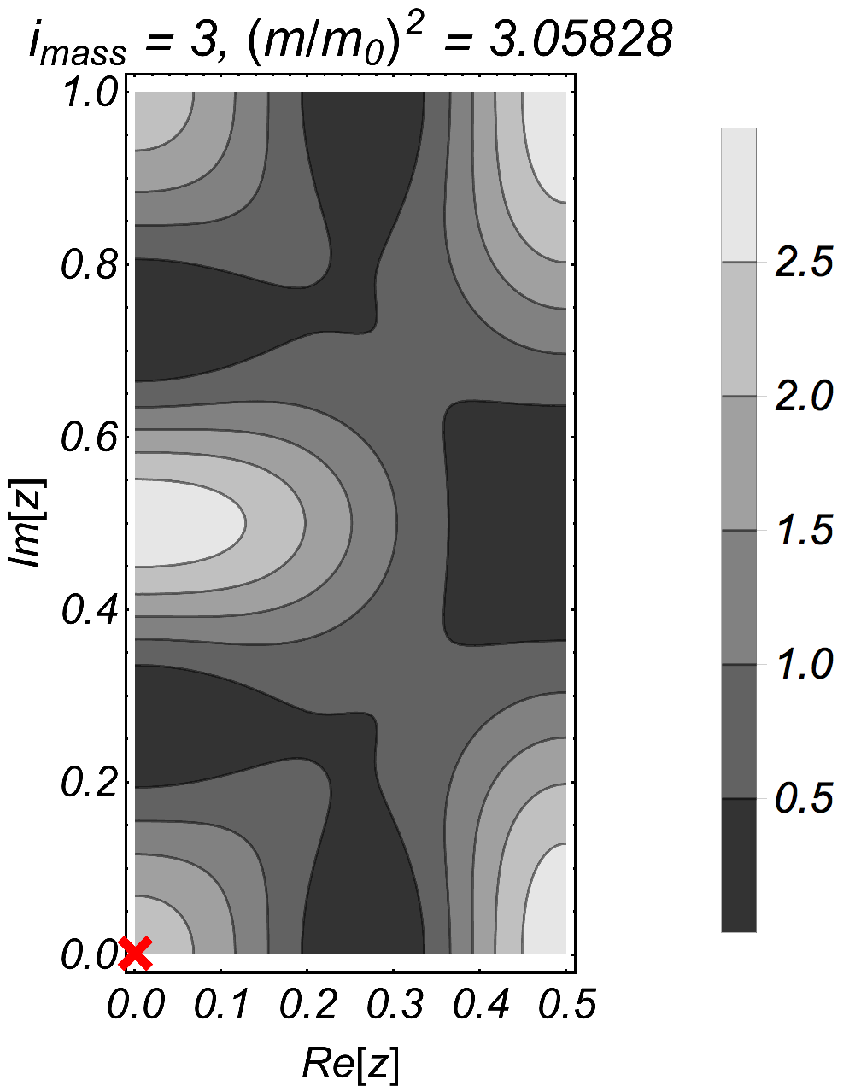}
\caption{The probability densities $|\phi^{i_{\rm mass}}(z)|^2$ of the KK \cred{wave functions} {with \cred{mass eigenvalues that are} unaffected ($i_\text{mass} = 1$)/perturbed ($i_\text{mass} = 2,\,3$)} \cred{by} the brane-localized mass at {$z = z_1$} for $M=3$ and $h=0.5$. A \cred{red cross} denotes a position with the brane-localized mass.
The corresponding mass eigenvalues after the perturbation are also shown.
{The values of the ratio $(m/m_0)^2$ before the perturbation are $1$, $1$\cred{,} and $3$, respectively.}}
\label{wavefuncs}
\end{figure}

\begin{figure}[H]
\centering
\includegraphics[width=0.3\textwidth]{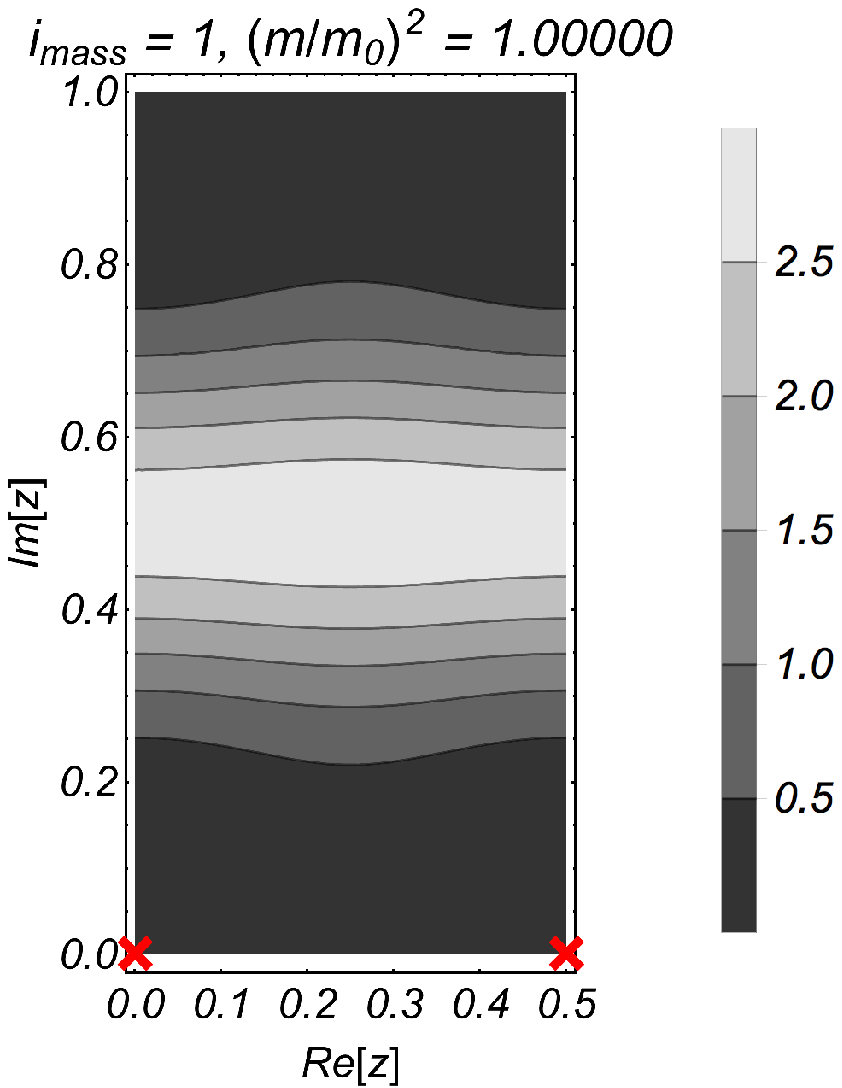}\hspace{0.1\textwidth}
\includegraphics[width=0.3\textwidth]{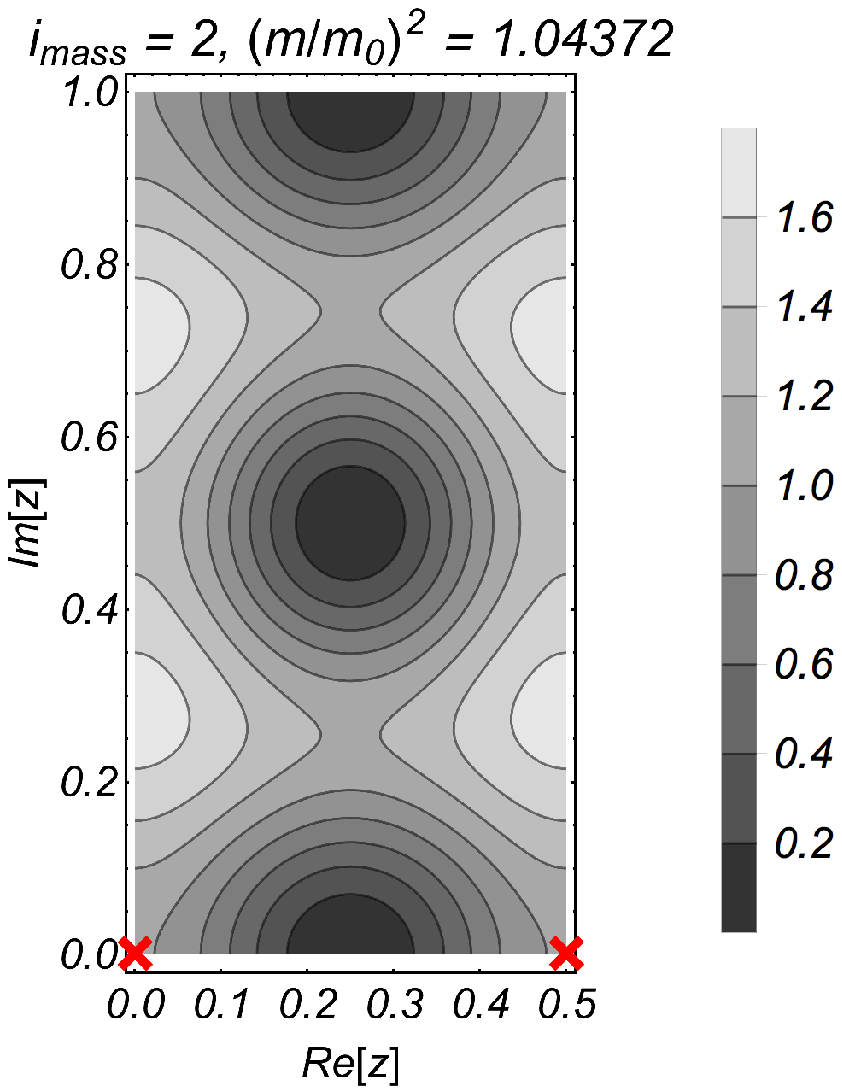}\\[10pt]
\includegraphics[width=0.3\textwidth]{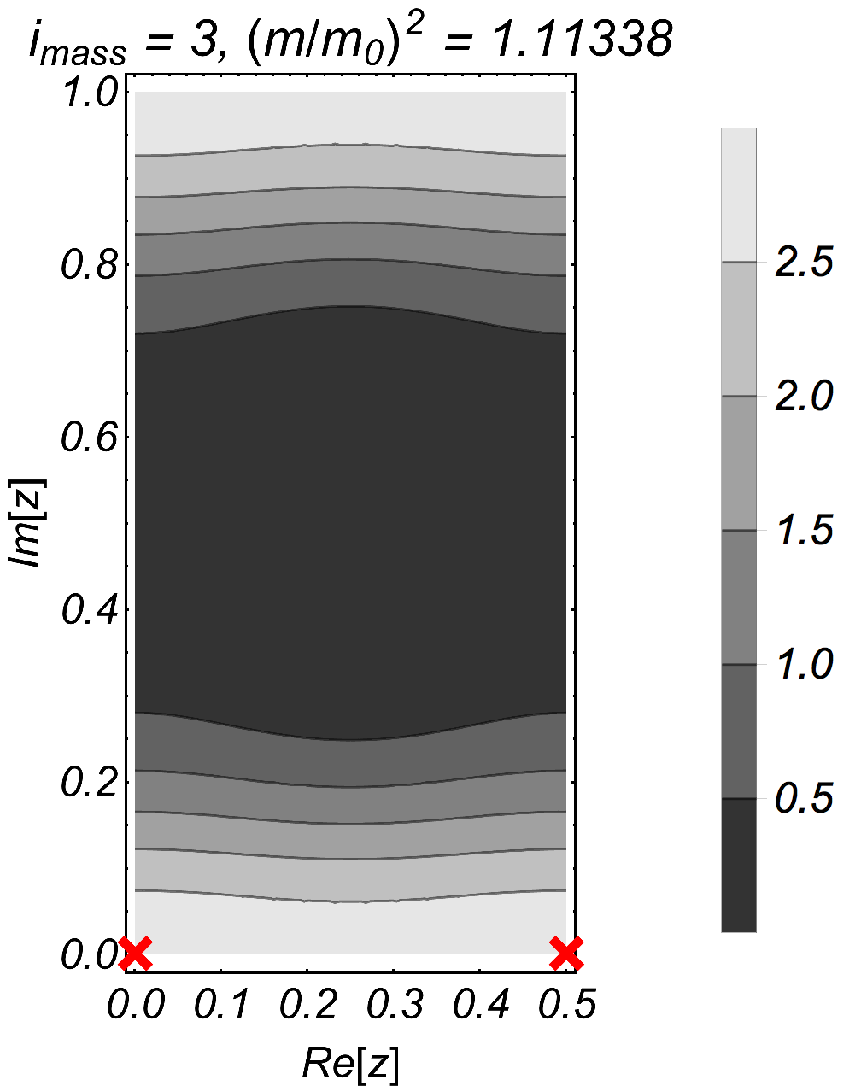}\hspace{0.1\textwidth}
\includegraphics[width=0.3\textwidth]{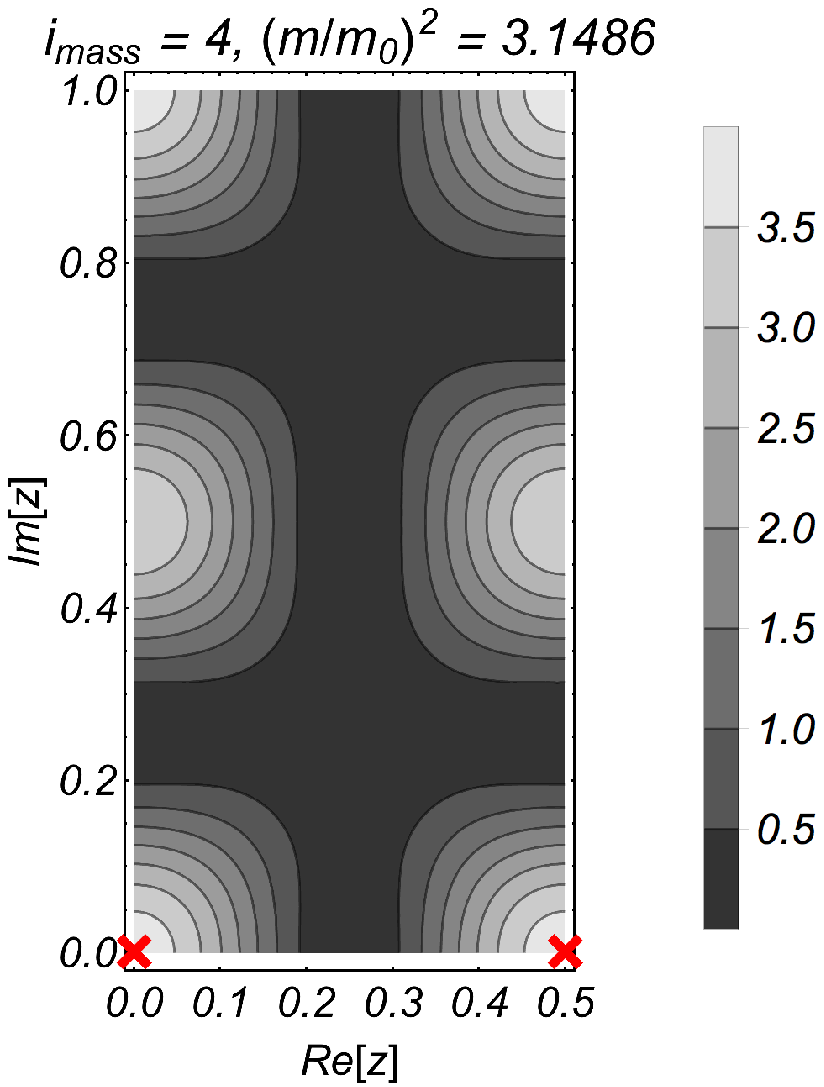}
\caption{The probability densities $|\phi^{i_{\rm mass}}(z)|^2$ of the KK \cred{wave functions} {with \cred{mass eigenvalues that are} unaffected ($i_\text{mass} = 1$)/perturbed ($i_\text{mass} = 2,\,3,\,4$) \cred{by} the brane-localized mass terms at {$z = z_1$} and {$z = z_2$}} for {$M=4$} and $h=0.5$.
\cred{The two red crosses} in each panel denote the positions where the brane-localized mass terms are located.
The corresponding mass eigenvalues after the perturbation are also shown.
{The values of the ratio $(m/m_0)^2$ before the perturbation are $1$, $1$, $1$\cred{,} and $3$, respectively.}}
\label{wavefuncs2}
\end{figure}

\cred{As} an {illustrative example}, we consider a simple example {in the scalar case} for $M=3$ and $\Lambda=m_1=18\pi/{\cal A}$, and then $\alpha=1, 2, 3$ \cred{[}corresponding to $(n,j)=(0,0), (0,1), (1,1)$\cred{]}.
Then, the KK mass matrix under consideration is the same as {the expression in} Eq.\,\eqref{3x3scalar}.
For $h=0.5$, the eigenvalues (divided by $m_0$) of Eq.\,\eqref{3x3scalar} are numerically obtained as
\begin{gather}
{(m/m_0)^2 = 1.00000, \ 1.08008, \ 3.05828},
\end{gather}
and also the corresponding eigenvectors are given as
\begin{gather}
{u^{(1)}} = (0.459701, -0.888074, 0), \
{u^{(2)}  = (-0.88755, -0.45943, 0.0343449)}, \notag \\
{u^{(3)}  = (0.0305008, 0.0157884, 0.99941)}.
\end{gather}
The corresponding \cred{wave functions} unaffected/affected by the brane-localized mass are obtained by internal products of the above eigenvectors and the KK \cred{wave functions} {before the perturbation} $\phi_\alpha \equiv (\phi^0_{T^2/Z_2, 0}, \phi^1_{T^2/Z_2, 0}, \phi^1_{T^2/Z_2, 1})$, i.e.,
\begin{gather}
\phi^{i_{\rm mass}}(z) \equiv \sum_{\alpha=1}^3 \left(u^{(i_{\rm mass})}\right)_\alpha \phi_\alpha (z)
\qquad
\left( i_\text{mass} = 1,\,2,\,3 \right).
\end{gather}
We show the probability densities $|\phi^{i_{\rm mass}}(z)|^2$ of the KK \cred{wave functions with mass eigenvalues that are} unaffected/perturbed \cred{by} the brane-localized mass at {$z = z_1$} for $M=3$ in Fig.\,\ref{wavefuncs}, where a \cred{red cross} denotes a position with the brane-localized mass.
Figure\,\ref{wavefuncs} tells that an unaffected mode ($i_{\rm mass}=1$) avoids the position with the localized mass, and also that the other affected modes ($i_{\rm mass}=2, 3$) are localized at the position with the localized mass. 
The {trend} that unaffected modes avoid the position with the localized mass is {also found in situations} with multiple localized masses {(see Fig.\,\ref{wavefuncs2})}.

We provide another example \cred{with} $M=4$ and $\Lambda=m_1$ ($\alpha=1,2,3,4$), and two brane-localized mass terms at the two fixed points $z=z_1$ and $z=z_2$.
The sketches of \cred{wave function} localizations are shown in Fig.\,\ref{wavefuncs2}.
Here, we can see that two of the three lowest modes before the perturbation (see \cred{Table} \ref{numofneven}) are uplifted.

\section{Cutoff dependence of mass eigenvalues}
\label{sec:cutoff}

Once we specify a cutoff scale $\Lambda$, we can write down the KK mass spectra with the brane-localized mass below $\Lambda$.
As concretely addressed in the previous section, {when $\eta = +1$ and $M \geq 2$, a part of} the lowest modes of \cred{the} KK mass spectrum in the scalar is unaffected \cred{by} the presence of a single localized mass term, where the \cred{values} of such unperturbed mass eigenstates \cred{are} independent of the cutoff scale $\Lambda$.
{On the other hand}, \cred{some} mass eigenvalues are perturbed and get heavier by the effect of the localized mass term, where the degree of such deformations would depend on the cutoff scale.
We suppose that $\Lambda = m_{n_{\rm max}}$ for a certain $n_{\rm max}$.
Hereafter, we consider $n_{\rm max}$ instead of $\Lambda$.

As \cred{before}, we focus on $M=3$.
The two lowest modes without localized masses are doubly degenerate and their KK masses \cred{are} ${m_0^2} = 6\pi/{\cal A}$.
After introducing a localized mass \cred{at} $z=z_1$, one of the two lowest modes gets heavier and {the perturbed mass} is roughly estimated as ${m^2 \sim m_0^2} (1 + {\cal O}(h))$, \cred{assuming that} $|v_\alpha R|^2 \sim {{\cal O}(1)}$.
For \cred{the maximum} of the KK levels $n_{\rm max}$, \cred{the} ratio {$(m/m_0)^2$} can be {interpolated} as
\begin{gather}
{\left(\frac{m}{m_0}\right)^2} \simeq \alpha_1 + \alpha_2 \ln n_{\rm max} + \alpha_3 \, n_{\rm max},
\end{gather}
where $\alpha_i \,\, (i=1,2,3)$ are real constants.
The cutoff scale dependence of {the (squared) mass ratio $(m/m_0)^2$} is shown in Fig.\,\ref{cutoffdep1} (Fig.\,\ref{cutoffdep2}), where we set $h=0.5$ ($h=3.0$), respectively.
Figures~\ref{cutoffdep1} and \ref{cutoffdep2} tell us that \cred{the choic of} cutoff scale does not drastically \cred{affect} the mass ratio, and we can conclude that such a cutoff dependence is {irrelevant} {from} {a \cred{model-building} point of view} \cred{[}{except for $h$ being \cred{as} huge as $\gtrsim \mathcal{O}(10)$ and/or the compactification scale $\sim R^{-1}$ being \cred{as} small as $\lesssim {\mathcal{O}(1)}\,\text{TeV}$}\cred{]}.
We also find that the cutoff dependence {looks} similar and \cred{would} be irrelevant for greater magnitudes of the fluxes.\footnote{
{Here, we comment on approaches to treat \cred{brane-localized} mass terms.
The \cred{simplest} method adopted in our \cred{analysis---where} an {effective} mass matrix with infinite numbers of columns and lows is derived through KK expansions, and we (numerically) diagonalize an approximated form with a truncation of higher modes holding heavier KK masses than a cutoff scale $\Lambda$\cred{---is} enough when results are not sensitive to values of the cutoff.
For more precise discussions, the techniques with the theta functions argued in Refs.~\cite{Dudas:2005vn,Buchmuller:2016gib} would be useful.
See also\cred{,} e.g., Refs.~\cite{Diener:2013xpa,Barcelo:2014kha} (and references therein) for discussions on \cred{the} bulk-boundary interplay of higher-dimensional fields.}
}
{Also, we would like to comment on the testability of the mass correction via localized masses through \cred{high-energy experiments}.
When the coefficient of a localized mass \cred{$h$} is small, the cutoff dependence is negligible \cred{(}as already explained\cred{) and it seems} difficult to probe the effect by discovering several KK modes and \cred{measuring} the mass differences among them.
Thus, the mechanism has difficulties from the testability point of view.
However, \cred{the main motivation} of this research is to reveal the relationship between the presence of \cred{localized} masses and the number of fluxes (corresponding to the number of unperturbed matter generations).

{When we consider a high-scale extra-dimensional theory, the correction in KK states (before the mass perturbation) seems to be less important.
Nevertheless, we can claim that the \cred{stability} of {the values of heavier states} (compared with the electroweak scale) \cred{against} the perturbation is a good \cred{feature of} the present scenario.}

\begin{figure}[H]
\centering
\includegraphics[width=0.55\textwidth]{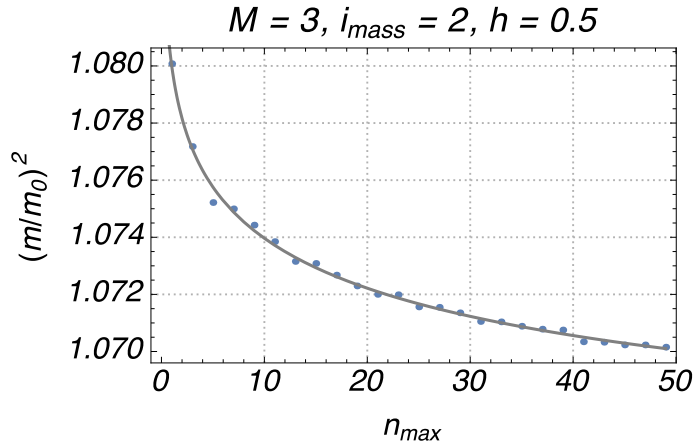}
\caption{The cutoff scale dependence of the squared mass ratio $(m/m_0)^2$ {for a perturbed mode}.
Fitting parameters are chosen as $(\alpha_1, \alpha_2, \alpha_3) = {(1.07999, -0.00265445, 9.10443 \times 10^{-6})}$.
{Note that the ratio before the perturbation is $(m/m_0)^2 = 1.$}}
\label{cutoffdep1}
\end{figure}

\begin{figure}[H]
\centering
\includegraphics[width=0.55\textwidth]{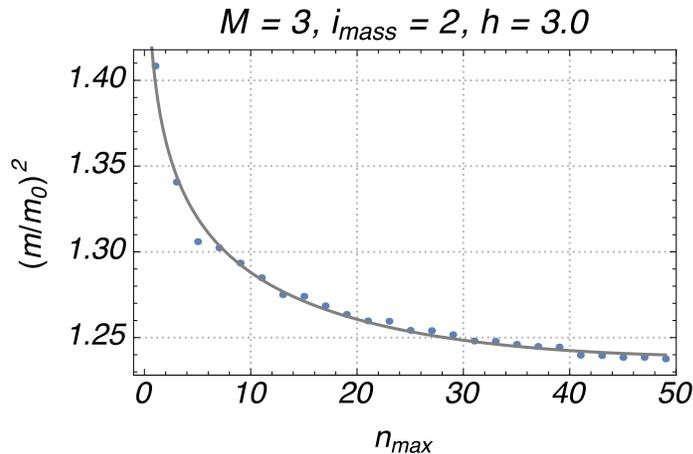}
\caption{The cutoff scale dependence of the squared mass ratio $(m/m_0)^2$ {for a perturbed mode}.
Fitting parameters are chosen as $(\alpha_1, \alpha_2, \alpha_3) = {(1.39949, -0.0523681, 0.0009029)}$.
{Note that the ratio before the perturbation is $(m/m_0)^2 = 1.$}}
\label{cutoffdep2}
\end{figure}

\section{Conclusion}
\label{sec:conclusion}

In this paper, we have considered effects on the \cred{KK} mass spectra via the presence of brane-localized masses \cred{at} fixed points of a toroidal orbifold $T^2/Z_2$ with magnetic fluxes.
Under the presence of the magnetic fluxes on the toroidal and orbifold compactifications, the magnetic fluxes {are quantized and become topological indices}, and then the multiplicity of KK-decomposing \cred{wave functions} appears in the {low-energy} effective theory.

We have added \cred{single} or multiple brane-localized masses \cred{at} the fixed points {of $T^2/Z_2$ where some {parts} of \cred{the} KK spectrum on $T^2$ are projected out {by the orbifolding}, {while} multiple degenerated modes {still} \cred{remain} (if $M \geq 2$)}.
By analyzing the effects of the localized masses \cred{using} linear algebra, we have found that, \cred{at} each KK level, \cred{one} or more \cred{of} the degenerate KK masses are perturbed, when {single or multiple brane-localized mass terms are introduced}.
This discussion is valid for both of the {six-dimensional} scalar and spinor fields.
In addition, we have also investigated deformation in \cred{wave functions} through the localized mass terms and the cutoff dependence of the magnitude of modulations in \cred{the} mass eigenvalues.

The mechanism which we have investigated \cred{in} this paper is useful for phenomenologies on magnetized orbifolds, especially {in constructing unified theories with a much heavier KK scale compared with the electroweak scale}, for decoupling some light exotic particles away from the physics around the electroweak and \cred{TeV} scales.
{An important point in the mechanism in 6D is that particle spectra do not seriously depend on the magnitude of \cred{the} coefficients of brane-localized mass terms when the KK scale is sufficiently higher than the electroweak scale, while the {number} of \cred{fixed points} where brane-localized mass terms are injected {plays} a significant role.
\cred{Therefore}, we can conclude that our mechanism is useful for removing unwanted exotic light modes from low-energy effective theories without relying on \cred{the} details of theories in extra dimensions.}

Applications to more generalized situations with nontrivial \cred{Scherk-Schwarz} and Wilson line phases as well as general choices in the complex structure parameter $\tau$ look fruitful.

\section*{\cred{Acknowledgments}}
M.I. and Y.T. would like to thank Hiroyuki Abe for helpful discussions.
K.N. is grateful to Hiroyuki Abe and the
\cred{particle physics group at} Waseda University for the kind hospitality \cred{during} the final stage of this work.
{We thank the \cred{PRD referee} for giving us various fruitful comments.}
Y.T. is supported in part by Grants-in-Aid for Scientific Research No.\,16J04612 from the Ministry of Education, Culture, Sports, Science and Technology \cred{of} Japan.

\appendix
\renewcommand{\thesection}{\Alph{section}}

\section{Notation}\label{app:notation}

{In this appendix, we review \cred{our notation for} 6D gamma matrices, which obey the Clifford algebra
\begin{align}
\{ \Gamma^M, \Gamma^N \} = -2 g^{MN},\qquad
g^{MN} = \text{diag}(-1,+1,+1,+1,+1,+1).
\end{align}
Our choice \cred{for} the set of 6D gamma matrices is as follows\cred{:}
\begin{align}
\Gamma^\mu =
	\begin{pmatrix}
	\gamma^\mu & 0 \\ 0 & \gamma^\mu
	\end{pmatrix},
\qquad
\Gamma^{5} =
	\begin{pmatrix}
	0 & i\gamma_5  \\ i\gamma_5 & 0
	\end{pmatrix},
\qquad
\Gamma^{6} =
	\begin{pmatrix}
	0 & \gamma_5 \\ -\gamma_5 & 0
	\end{pmatrix},
\end{align}
where $\gamma_5$ describes the 4D chirality, which is defined as $\gamma_5 = i \gamma^0 \gamma^1 \gamma^2 \gamma^3$.
The matrix denotes the 6D chirality and can be decomposed as
\begin{align}
\Gamma^7 = - \Gamma^0 \Gamma^1 \Gamma^2 \Gamma^3 \Gamma^5 \Gamma^6
=
	\begin{pmatrix}
	\gamma_5 & 0 \\ 0 & -\gamma_5
	\end{pmatrix}
= \Gamma_{\text{4D chiral}} \, \Gamma_{\text{internal}}
\end{align}
with
\begin{align}
\Gamma_{\text{4D chiral}} = 
	\begin{pmatrix}
	\gamma_5 & 0 \\ 0 & \gamma_5
	\end{pmatrix},\qquad
\Gamma_{\text{internal}} \equiv i \Gamma^{5} \Gamma^{6} =
	\begin{pmatrix}
	I_4 & 0 \\ 0 & -I_4
	\end{pmatrix}.
\end{align}
The matrix $\Gamma_{\text{internal}}$ describes eigenvalues of the internal chirality.
The 6D chirality is calculated \cred{as the simple product of the} 4D chirality and the internal chirality\cred{.}}

\section{Generic discussion in the scalar case}\label{app:multiple_massterm}

In the main sections, we mainly \cred{discussed the effects arising from} the presence of a single brane-localized mass.
In this appendix, we extend the discussion to multiple brane-localized mass terms and their effects on KK mass eigenvalues.
The scalar case is \cred{addressed} in the following discussion, where \cred{the method} to treat the fermion case is straightforwardly recognized through the result \cred{for} the scalar.
The reason is that $\mathcal{M}^2$ for the scalar and $\mathcal{M}^\dagger \mathcal{M}$ for the fermion \cred{take} the same form, as we pointed out in \cblue{{Sec.}\,\ref{sec:spinor}}.

For example, we add another brane-localized mass to Eq.\,\eqref{symb}, and then obtain
\begin{gather}
{\cal M}^2 = {\rm diag} \, (\underbrace{m_0^2, m_0^2, ..., m_0^2}_{d_e}, \, 
                            \underbrace{m_1^2, m_1^2, ..., m_1^2}_{d_o}, \, 
                            \underbrace{m_2^2, m_2^2, ..., m_2^2}_{d_e}, \cdots ) + 
                            (v^{(1)})^{\dag} \otimes v^{(1)} + (v^{(2)})^{\dag} \otimes v^{(2)}.
\end{gather}
Here, we do not specify two distinct positions with the localized masses and symbolically express $(v^{(1)})_\alpha = \phi_\alpha(z_i)$ and $(v^{(2)})_\alpha = \phi_\alpha(z_j)$ for $z_i \neq z_j$.
In order to keep one of the original lowest {eigenvalues} as $m_0^2$ after diagonalizing the KK mass matrix, an eigenvector $(u)_\alpha$ which \cred{contains}
\begin{gather}
(u)_\alpha = (u_1, u_2, ..., u_{d_e}, {0,0,\cdots})
\end{gather}
{should simultaneously satisfy}
\begin{align}
v^{(1)} \cdot u = 0 \ &\iff \ v^{(1)}_1 u_1 + v^{(1)}_2 u_2 + ... + v^{(1)}_{d_e} u_{d_e} = 0, \label{internalprod1}\\
v^{(2)} \cdot u = 0 \ &\iff \ v^{(2)}_1 u_1 + v^{(2)}_2 u_2 + ... + v^{(2)}_{d_e} u_{d_e} = 0. \label{internalprod2}
\end{align}
{The relations in Eqs.\,(\ref{internalprod1}) and (\ref{internalprod2}) ensure that $(u)_\alpha$ is} an eigenvector with the eigenvalue $m_0^2$,
\begin{align}
\mathcal{M}^2 \, u = m_0^2 \, u + 
		(v^{(1)})^{\dag} \otimes \left( v^{(1)} \cdot u \right) + 
		(v^{(2)})^{\dag} \otimes \left( v^{(2)} \cdot u \right) =
		m_0^2 \, u.
\end{align}

When we recognize that whether an eigenvector is normalized or not does not affect the number of linearly independent eigenvectors, {we find that} at least three nonzero components are required in $u$ as,
\begin{align}
(u^{(1)})_\alpha = (u_1^{(1)},u_2^{(1)},u_3^{(1)},{0,0,}\cdots),
\end{align}
which {obeys} the simplified constraints,
\begin{align}
v^{(1)}_1 u^{(1)}_1 + v^{(1)}_2 u^{(1)}_2 + v^{(1)}_3 u^{(1)}_3  &= 0, \\
v^{(2)}_1 u^{(1)}_1 + v^{(2)}_2 u^{(1)}_2 + {v^{(2)}_3} u^{(1)}_3  &= 0. 
\end{align}
When we take $u^{(1)}_1 = 1$, which is just a scaling, the corresponding \cred{values} of $u^{(1)}_2$ and $u^{(1)}_3$ are fixed as
\begin{align}
u^{(1)}_2 = \frac{v^{(1)}_3 v^{(2)}_1 - v^{(1)}_1 v^{(2)}_3}{-v^{(1)}_3 v^{(2)}_2 + v^{(1)}_2 v^{(2)}_3},\qquad
u^{(1)}_3 = \frac{v^{(1)}_2 v^{(2)}_1 - v^{(1)}_1 v^{(2)}_2}{ v^{(1)}_3 v^{(2)}_2 - v^{(1)}_2 v^{(2)}_3}.
\end{align}

Apparently, similar procedures can continue, e.g., for $(u^{(2)})_\alpha = (0,u_2^{(2)},u_3^{(2)},u_4^{(2)},{0,0,}\cdots)$.
Now, we can conclude that the number of linearly independent eigenvectors under two brane-localized mass terms is $d_e - 2$ (for even $n$) and $d_o -2$ (for odd $n$), respectively, unless anomalous situations \cred{arise}, e.g., $-v^{(1)}_3 v^{(2)}_2 + v^{(1)}_2 v^{(2)}_3 = 0$.
In such exceptionally special {cases}, the number of linearly independent eigenvectors does not obey the above criterion.

{Situations with brane-localized mass terms \cred{at} three (four) fixed points are scrutinized in the same way,
where $d_e - 3$ ($d_e - 4$) [for even $n$] and $d_o -3$ ($d_o -4$) [for odd $n$] independent physical modes remain \cred{unperturbed} in {\cred{the} case without} accidental cancellation in corresponding conditions.}


\bibliographystyle{utphys}
\bibliography{references_ver2}
\end{document}